\documentstyle[epsf,aps]{revtex}
\renewcommand{\thefootnote}{\fnsymbol{footnote}}
\begin{document}
\newcommand{\be}{\begin{eqnarray}}
\newcommand{\dlq}{\lq\lq}
\newcommand{\ee}{\end{eqnarray}}
\newcommand{\ben}{\begin{eqnarray*}}
\newcommand{\een}{\end{eqnarray*}}
\newcommand{\stackeven}[2]{{{}_{\displaystyle{#1}}\atop\displaystyle{#2}}}
\newcommand{\lsim}{\stackeven{<}{\sim}}
\newcommand{\gsim}{\stackeven{>}{\sim}}
\newcommand{\un}[1]{\underline{#1}}
\renewcommand{\baselinestretch}{1.0}
\newcommand{\as}{\alpha_s}
\newcommand{\tas}{{\tilde\alpha}_s}
\newcommand{\bas}{{\overline\alpha}_s}
\def\eq#1{{Eq.~(\ref{#1})}}
\def\fig#1{{Fig.~\ref{#1}}}
\begin{flushright}
BNL-NT-03/14 \\
NT@UW--03--016 \\
INT-PUB--03--13
\end{flushright}
\vspace*{1cm} 
\setcounter{footnote}{1}
\begin{center}
{\Large\bf Cronin Effect and High-$p_T$ Suppression in pA Collisions}
\\[1cm]
Dmitri Kharzeev$^{1}$, Yuri V.\ Kovchegov$^{2}$ and Kirill Tuchin$^{3}$ 
\\ ~~ \\ 
{\it $^1$ Nuclear Theory Group, Brookhaven National Laboratory, Bldg. 510A} \\ 
{\it Upton, NY 11973} \\ ~~ \\ 
{\it $^2$ Department of Physics, University of Washington, Box 351560} \\
{\it Seattle, WA 98195 } \\ ~~ \\ 
{\it $^3$ Institute for Nuclear Theory, University of Washington, Box 351550} 
\\ {\it Seattle, WA 98195 } \\ ~~ \\ ~~ \\
\end{center}
\begin{abstract}
We review the predictions of the theory of Color Glass Condensate for
gluon production cross section in p(d)A collisions. We demonstrate
that at moderate energies, when the gluon production cross section can
be calculated in the framework of McLerran-Venugopalan model, it has
only partonic level Cronin effect in it. At higher energies/rapidities
corresponding to smaller values of Bjorken $x$ quantum evolution
becomes important. The effect of quantum evolution at higher
energies/rapidities is to introduce suppression of high-$p_T$ gluons
slightly decreasing the Cronin enhancement. At still higher
energies/rapidities quantum evolution leads to suppression of produced
gluons at all values of $p_T$.
\end{abstract}

\renewcommand{\thefootnote}{\arabic{footnote}}
\setcounter{footnote}{0}

\section{Introduction}

Recently there has been a surge of interest in particle production in
proton-nucleus (pA) and deuteron-nucleus (dA) collisions at high
energies. The interest was inspired by the new data produced by the dA
program at RHIC \cite{dAtaphen,dAtaphob,dAtastar,Arsene:2003yk}, which
should enable one to separate the contributions of the initial state
effects \cite{KLM} such as parton saturation \cite{glr,mq,BM,mv,kjklw}
from the final state effects, such as jet quenching and energy loss in
the quark-gluon plasma (QGP) \cite{Bj,EL,BDMPS,EL2}, to the
suppression of high transverse momentum particles observed in $Au-Au$
collisions at RHIC \cite{aaphenix,aaphobos,aastar}.

Saturation physics has been largely successful in describing hadron
multiplicities in Au-Au collisions at RHIC \cite{KLN}. It can also
have important implications for the transverse momentum distributions
\cite{SKMV} and particle correlations and azimuthal anisotropies
\cite{KTf}. It has been demonstrated \cite{BMSS} that saturation
provides very favorable initial conditions for the thermalization of
parton modes with the transverse momenta $k_T \sim Q_s$, where $Q_s$
is the saturation scale. The thermalization was also found \cite{BMSS}
to approximately preserve the centrality dependence of total hadron
multiplicities determined by the initial conditions \cite{KLN}. Recent
lattice results \cite{KV,lappi} show that the initial average
transverse momentum $\langle k_T \rangle$ of the produced partons is $
\langle k_T \rangle \sim Q_s$, which makes the ``soft thermalization''
scenario preserving the initial centrality and rapidity distributions
quite likely.  Final state interactions, however, will undoubtedly
modify the transverse momentum distributions at $k_T \leq (1 \div 3)\
Q_s$ without introducing new momentum scale \cite{BMSS}. If the
produced medium lives long enough, then high $k_T$ jets will be
suppressed as well because of the jet quenching and energy loss
\cite{Bj,EL,BDMPS,EL2}.

The first $d-Au$ data from RHIC show Cronin enhancement extending up
to $k_T \simeq 6$ GeV around $y \sim 0$
\cite{dAtaphen,dAtastar} whereas at slightly forward rapidity around 
$y \sim 1$ no significant enhancement is seen \cite{dAtaphob}. The
absence of suppression indicates that final state interactions are
indeed responsible for the effect observed in $Au-Au$ collisions
\cite{aaphenix,aaphobos,aastar} in the same $k_T$ range. However, the
non--universality of the ratios for the charged hadron and $\pi^0$
spectra \cite{dAtaphen} indicate deviations from the independent jet
fragmentation up to $k_T \simeq 5$ GeV. Similar non-universality in
the same $k_T$ range was observed for $\Lambda$ and $K$ production
\cite{binstar}, and in $p, \bar{p}$ and pion production
\cite{Adler:2003qs} in $Au-Au$ collisions. It remains to be checked if
there is a statistically significant suppression of high $k_T$ charged
hadron and $\pi^0$ yields above the Cronin enhancement region ($k_T
\geq 6$ GeV), and if this suppression depends on centrality of $d-Au$
collisions. This question is of crucial importance for the
interpretation of the spectacular effect observed in $Au-Au$
collisions \cite{aaphenix,aaphobos,aastar} because this is the
kinematical region in which the independent jet fragmentation picture,
and thus the perturbative jet quenching description, begin to apply.

The first dA data from RHIC \cite{dAtaphen,dAtaphob,dAtastar} give the
ratio of the number of particles produced in a dA collision over the
number of particles produced in a pp collision scaled by the number of
collisions
\be\label{rda}
R^{dA} (k_T, y) \, = \, \frac{\frac{d N^{dA}}{d^2 k \ dy}}{N_{coll}
\, \frac{d N^{pp}}{d^2 k \ dy}}.
\ee
To understand the new data on $R^{dA}$ and what it implies for our
understanding of high energy nuclear wave functions we are going to
study here the expectations for $R^{dA}$ from saturation physics. Our
approach will be somewhat academic: in this paper, we will not include
explicitly all of the effects related to the fact that high-$k_T$ of
produced particles corresponds to rather large Bjorken $x$ in the
actual RHIC experiments at central rapidity -- the effective Bjorken
$x$ of high-$k_T$ ($k_T > 5$~GeV) particles observed at mid-rapidity
at RHIC at $\sqrt{s} = 200$~GeV is about $x \approx 0.1$ which may be
too large for the small-$x$ treatment that we present here (see
\cite{KS}, but see also \cite{GLMMT}). These finite--energy 
effects have to be accurately accounted for before we can compare our
calculations to the data. Nevertheless, we feel that a better
understanding of the qualitative features of hadron production within
the saturation framework is a necessary pre--requisite for a complete
theoretical description of high energy $p(d)-A$ collisions.

We assume that collisions take place at very high energy such that the
effective Bjorken $x$ is sufficiently small for all $k_T$ of
interest. For simplicity we will analyze proton-nucleus collisions
assuming that the main qualitative conclusions would be applicable to
dA. Since we can not calculate $N_{coll}$ in a model-independent way,
we will be using \eq{rpa} for our definition of $R^{pA}$, which is
identical to \eq{rda} applied to pA collisions with a proper
definition of $N_{coll}$ (see \cite{Boris} for a discussion of
uncertainties involved in theoretical evaluations of this quantity).

The problem of gluon production in $pA$ collisions has been solved in
the framework of McLerran-Venugopalan model \cite{mv} in \cite{KM}
(see also \cite{KTS,DM,KW1,yuriaa}). The resulting cross section
includes the effects of all multiple rescatterings of the produced
gluon and the proton in the target nucleus \cite{KM}. At higher energy
quantum evolution becomes important
\cite{BFKL,yuri1,yuri2,dip,bal,JKLW,FILM,Braun1}. In the large $N_c$ limit 
the small-$x$ evolution equation can be written in a non-linear
integro-differential form \cite{yuri1,yuri2,dip,bal} shown here in
\eq{eqN}. The inclusion of non-linear evolution
\cite{yuri1,yuri2,dip,bal} in the quasi-classical gluon production
cross section of \cite{KM} has been done in \cite{KT} (see also
\cite{Braun2,yuridiff}). The study of the resulting gluon spectrum and
corresponding gluonic $R^{pA}$ is the goal of this paper.

The paper is organized as follows. In Sect. IIA we discuss two main
definitions of unintegrated gluon distribution functions: the standard
definition (\ref{ktglue}) and the one inspired by non-Abelian
Weizs\"{a}cker-Williams field of a nucleus in McLerran-Venugopalan
model (\ref{wwglue}) \cite{mv,kjklw}. We argue, following
\cite{KT,IIM1}, that the \eq{wwglue} is the correct definition of the
unintegrated gluon distribution counting the number of gluon
quanta. We proceed by analyzing $k_T$-dependence of the distribution
functions. In Sect. IIB we prove the sum rules for both distribution
functions given in Eqs. (\ref{sumkt}) and (\ref{sumww}), which are
valid in the quasi-classical approximation only. In the framework of
McLerran-Venugopalan model \cite{mv,Mueller1} the sum rules insure
that the presence of shadowing in nuclear gluon distribution functions
in the saturation regime ($k_T \lsim Q_{s0}$) requires enhancement of
gluons at higher $k_T$ ($k_T \gsim Q_{s0}$) reminiscent of
anti-shadowing.  This conclusion is quantified in Sect. IIC [see
\fig{shad} and Eqs. (\ref{wwra}) and (\ref{ktra})] and the differences
between distribution functions are clarified [see Eqs. (\ref{ktsat})
and (\ref{wwsat})]. However, as we demonstrate in Sect. IIB, the sum
rules break down once quantum evolution with energy
\cite{yuri1,yuri2,dip,bal} is included. They turn into inequalities 
(\ref{inkt}) and (\ref{inww}). This indicates that, while multiple
rescatterings in McLerran-Venugopalan model only redistribute gluons
in transverse momentum phase space conserving the total number of
gluons in the nucleus \cite{KM}, quantum evolution of
\cite{yuri1,yuri2,dip,bal} actually reduces the number of gluons
in the nuclear wave functions at a given value of Bjorken $x$.

In Sect. III we study the gluon production cross section in pA in the
quasi-classical approximation \cite{KM}. In Sect. IIIA we show that
the gluon production cross section calculated in \cite{KM} in
McLerran-Venugopalan multiple rescattering model exhibits only
Cronin-like enhancement \cite{Cronin}, as shown in \fig{cron} and in
\eq{qclt} (cf. \cite{KNST,AG}).  In the corresponding moderately 
high energy regime the height and $k_T$-position of the Cronin peak
are increasing functions of centrality as can be seen from
\eq{max}. In Sect. IIIB following \cite{KT} we point out that,
surprisingly, the gluon production cross section in pA can be written
in a $k_T$-factorized form (\ref{kt}) \cite{glr,Ryskin} with the
unintegrated distribution functions defined by \eq{ktglue}, physical
meaning of which is less clear than that of Weizs\"{a}cker-Williams
ones (\ref{wwglue}).  In Sect. IIIB we also prove a sum rule
(\ref{sumr}) which insures that suppression of produced gluons at low
$k_T$ ($k_T \lsim Q_{s0}$) demands Cronin-like enhancement at high
$k_T$ ($k_T \gsim Q_{s0}$) in McLerran-Venugopalan model. The relative
amounts of suppression and enhancement are different from the
quasi-classical gluon distribution case of Sect. II.

Multiple rescatterings of partons inside the nucleus are believed to
be the cause of Cronin effect. Phenomenologically these multiple
rescatterings are usually modeled by introducing transverse momentum
broadening in the nuclear structure functions
\cite{ktbroadening1,ktbroadening2,Vitev03,ktbroadening3}. In Sect. III 
we demonstrate how an explicit pQCD calculation of these multiple
rescatterings done in \cite{KM} yields us Cronin effect
(cf. \cite{KNST,AG}).

Sect. IV is devoted to studying the effects of nonlinear evolution
(\ref{eqN}) on the gluon production cross section in pA. In Sect. IVA
we use the analogy to the case of gluon production in deep inelastic
scattering (DIS) solved in \cite{KT,yuridiff} to include the effects
of evolution (\ref{eqN}) in the gluon production cross section in
pA. The result is given by \eq{paev}. By expanding the all-twist
formula (\ref{paev}) we then study the effect of nonlinear evolution
on the gluon production at the leading twist (Sect. IVB) and
next-to-leading twist (Sect. IVC) level. In Sect. IVB we start by
deriving gluon production cross section at the leading twist level
(\ref{lt}). We then estimate the cross section for high $k_T$ ($k_T >
k_{\rm geom} \gg Q_s (y)$) in the double logarithmic approximation
(\ref{dla}) \cite{Ryskin} and demonstrate that the corresponding
$R^{pA}$ is approaching $1$ at high $k_T$ from below, i.e., that
$R^{pA} < 1$ at $k_T \gg Q_s (y)$ [see \eq{rxi1}]. We proceed by
evaluating \eq{lt} in the extended geometric scaling region ($Q_s (y)
< k_T \, \lsim \, k_{\rm geom}$) \cite{IIM1,geom,LT}. The resulting
leading twist gluon production cross section (\ref{lla}) leads to
further suppression of gluon production due to the change in gluon
anomalous dimension \cite{KLM} as shown in Eqs. (\ref{llar}) and
(\ref{rpae}). At very high energies when the gluon production in $pp$
is also in the extended geometric scaling region ($k_T < k_{\rm
geom}^p$) the ratio $R^{pA}$ saturates at $R^{pA} \sim A^{-1/6}$, as
follows from \eq{anas}. The next-to-leading twist contribution to the
gluon production cross section in pA is evaluated in Sect. IVC with
the result given by Eqs. (\ref{dla2}) and (\ref{lla2}). One can see
that the subleading twist term contributes towards enhancement of
gluon production at high $k_T$. However, in the $k_T$ region where the
next-to-leading twist contribution dominates over higher twists it is
small compared to the leading twist term. Therefore the positive sign
of the higher twist term can not alter our conclusion of high-$k_T$
suppression we derived by analyzing leading twist. To understand how
all twists add up we study what happens to Cronin peak ($k_T \sim Q_s
(y)$) at high energy in Sect. IVD. We find that the height of the
Cronin maximum decreases with energy and eventually Cronin peak
flattens out at the same level as the rest of $R^{pA}$ at higher
$k_T$, which is shown in \eq{s3}. In Sect. IVE we observe that
inclusion of evolution only strengthens the suppression of $R^{pA}$ at
low $k_T$ ($k_T \ll Q_s (y)$) [see \eq{s7}] which was observed before
in Sect. IIIB in the quasi-classical case. In Sect. IVF we construct a
toy model illustrating the conclusion of Sect. IV that quantum
evolution \cite{yuri1,yuri2,dip,bal} introduces suppression of
$R^{pA}$ at all values of $k_T$ (see \fig{toy}). We demonstrate that
quantum evolution not only suppresses $R^{pA}$ making it less than
$1$, but also turns $R^{pA}$ into a decreasing function of collision
centrality contrary to the quasi-classical expectations.

We conclude in Sect. V by summarizing our results.


\section{A Tale of Two Gluon Distribution Functions}

\subsection{Definitions}

\begin{figure}
\begin{center}
\epsfxsize=5cm
\leavevmode
\hbox{ \epsffile{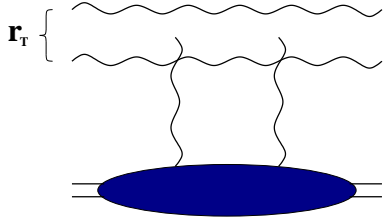}}
\end{center}
\caption{``Conventional'' definition of unintegrated gluon distribution 
relating it to the gluon dipole cross section. The exchanged gluon
lines can connect to either gluon in the dipole. }
\label{glue1}
\end{figure}

There are two different ways to define unintegrated gluon distribution
function of a proton or nucleus. The most conventional way relates it
to the $q\bar q$ dipole cross section on the target nucleus via two
gluon exchange. Here we are going to use a similar definition relating
the unintegrated gluon distribution to the dipole cross section on the
nucleus (see \fig{glue1}).
The corresponding gluon distribution is given by
(cf. \cite{Braun1,Braun2})
\be\label{ktglue}
\phi (x, \un k^2) \, = \, \frac{C_F}{\as \, (2 \pi)^3} \, \int d^2 b \, 
d^2 r \, e^{- i {\un k} \cdot {\un r}} \ \nabla^2_r \, N_G ({\un r},
{\un b}, y = \ln 1/x),
\ee
where $N_G ({\un r}, {\un b}, y = \ln 1/x)$ is the forward amplitude
of a gluon dipole of transverse size ${\un r}$ at impact parameter
${\un b}$ and rapidity $y$ scattering on a nucleus
\cite{yuri1,KT}. We denote by  $\un k$ the transverse
components of the four-vector $k$, and by $k_T$ its length.  The
definition of \eq{ktglue} is inspired by $k_T$-factorization and is
valid as long as one can neglect multiple rescatterings of the dipole
in the nucleus. By using \eq{ktglue} in the saturation region where
higher twists (multiple rescatterings) become important one implicitly
assumes that there exists a certain gauge in which the $q\bar q$
dipole cross section on a nucleus is given by a two gluon exchange
interaction between the dipole and the nucleus and the interaction
shown in \fig{glue1} is literally all one needs to obtain the correct
dipole cross section. It is not clear at present whether this is the
case and such a gauge exists. Therefore the gluon distribution given
by \eq{ktglue} does not give one the number of gluons in the nuclear
wave function in the saturation region. The applications of the
definition (\ref{ktglue}) will be clarified later.

\begin{figure}
\begin{center}
\epsfxsize=5cm
\leavevmode
\hbox{ \epsffile{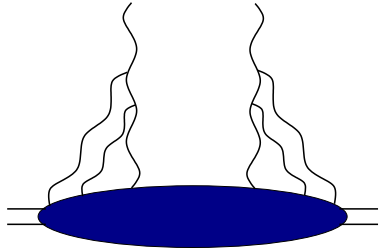}}
\end{center}
\caption{Definition of unintegrated gluon distribution in 
McLerran-Venugopalan model. }
\label{glue2}
\end{figure}

Another definition of unintegrated gluon distribution literally counts
the number of gluons in the nuclear wave function. To construct it in
the quasi-classical limit of high energy QCD given by
McLerran-Venugopalan model \cite{mv} one has to first find the
classical gluonic field of the nucleus in the light cone gauge of the
ultrarelativistic nucleus (non-Abelian Weizs\"{a}cker-Williams field)
and then calculate the correlator of two of such fields to get
unintegrated gluon distribution function (see \fig{glue2}).
The non-Abelian Weizs\"{a}cker-Williams field of a nucleus has been
found in \cite{kjklw}, leading to the following expression for the
corresponding gluon distribution \cite{kjklw,KM}
\be\label{ww1}
\phi^{WW} (x, \un k^2) \, &=& \, \frac{1}{2 \, \pi^2} \, \int d^2 b \, 
d^2 r \, e^{- i {\un k} \cdot {\un r}} \ \mbox{Tr} \left< {\un A}^{WW}
({\un 0}) \cdot {\un A}^{WW} ({\un r}) \right> \, \rangle\nonumber\\ &=& \, 
\frac{4 \, C_F}{\as \, (2 \pi)^3} \, \int d^2 b \, 
d^2 r \, e^{- i {\un k} \cdot {\un r}} \ \frac{1}{{\un r}^2} \ (1 -
e^{-{\un r}^2 Q_{s0}^2 \ln(1/r_T\Lambda) /4}),
\ee
where
\be\label{sat}
Q_{s0}^2({\un b})\,=\, 4\pi\, \as^2\,\rho\, T(\un b),
\ee
with $\rho$ the atomic number density in the nucleus with atomic
number $A$, $T(\un b)$ the nuclear profile function and $\Lambda$ some
infrared cutoff.

Generalizing \eq{ww1} to include non-linear small-$x$ evolution in it
\cite{yuri1,yuri2,dip,bal} is rather difficult. However, the problem of 
including small-$x$ evolution has been solved for $F_2$ structure
function and for the gluon production cross section in DIS
\cite{yuri1,KT}. Inspired by those examples we conjecture that replacing the
Glauber-Mueller \cite{Mueller1} forward gluon dipole amplitude on the
nucleus by its' fully evolved expression to be found from the
nonlinear evolution equation \cite{yuri1,bal}
\be
1 - e^{-{\un r}^2 Q_{s0}^2\ln(1/r_T\Lambda) /4} \,
 \, \, \Rightarrow \, \, \, N_G ({\un r},
{\un b}, y)
\ee
would give us the unintegrated gluon distribution function of a
nucleus in the general case:
\be\label{wwglue}
\phi^{WW} (x, \un k^2) \, = \, 
\frac{4 \, C_F}{\as \, (2 \pi)^3} \, \int d^2 b \, 
d^2 r \, e^{- i {\un k} \cdot {\un r}} \ \frac{1}{{\un r}^2} \ N_G ({\un r},
{\un b}, y = \ln 1/x).
\ee
Similar expression for gluon distribution was obtained earlier in
\cite{IIM1}.

An important observation concerning the two gluon distributions
presented above has been made in \cite{KT,Braun2}. It was shown that,
while the Weizs\"{a}cker-Williams gluon distribution of \eq{wwglue}
indeed has a clear physical meaning of counting the number of gluons
\cite{kjklw}, it is the gluon distribution inspired by $k_T$-factorization 
and given by \eq{ktglue} that enters gluon production cross section in
pA collisions and in DIS \cite{KT,Braun2}. More precisely, the gluon
production cross section including the effects of multiple
rescatterings and quantum evolution in it can be reduced to a
$k_T$-factorized form \cite{glr} with the unintegrated gluon
distribution of a nucleus given by \eq{ktglue} \cite{KT}. The authors
can not offer any simple physical explanation of this
paradox. Nevertheless we keep both distributions in the discussion
below keeping in mind that the first one is more relevant to particle
production in pA.


\subsection{$k_T$-dependence: General Arguments}

Both definitions of unintegrated gluon distribution (\ref{ktglue}) and
(\ref{wwglue}) have the same high-$k_T$ asymptotics in the
quasi-classical approximation (see e.g. \eq{ww1}), which reads
\be\label{asym}
\phi_A (x, \un k^2) \, = \, \phi^{WW}_A (x, \un k^2) \, = \, A \, 
\phi_N (x, \un k^2) 
\, = \, A \, \frac{\as \, C_F}{\pi} \, \frac{1}{\un k^2}, 
\hspace*{1cm} k_T \rightarrow \infty,
\ee
where the index $A$ ($N$) denotes gluon distribution in a nucleus
(nucleon). Therefore the distributions are equivalent at the level of
leading twist, i.e., as long as we include only a single rescattering
in the dipole amplitude $N_G$.

In the quasi-classical case of McLerran-Venugopalan model both gluon
distributions obey a sum rule which we are going to prove here for
$\phi$. From \eq{ktglue} one can easily infer that
\be\label{sum1}
\int d^2 k \, \phi_A (x, \un k^2) \, = \, \frac{C_F}{\as \, (2 \pi)} 
\, \int d^2 b \ \left( \nabla^2_r \ N_G ({\un r},
{\un b}, y = \ln 1/x) \right) \bigg|_{{\un r} = 0}.
\ee
At very small $r_T$ the dipole cross section $N_G$ in
McLerran-Venugopalan model goes to zero as $r_T^2$ (color transparency
\cite{FMS}) with the coefficient in the front proportional to
$A^{1/3}$. One can see that this is explicitly true for the
Glauber-Mueller expression for the dipole cross section $N_G$
\cite{Mueller1}
\be\label{GM}
N_G ({\un r}, {\un b}, y=0) \, = \, 1 - e^{-{\un r}^2 Q_{s0}^2 
\ln(1/r_T\Lambda)/4}.
\ee
For $N_G$ from \eq{GM} we observe that
\be\label{subtr}
\lim_{r_T \rightarrow 0} \left( \nabla^2_r \ N_G ({\un r},
{\un b}, y = 0) - A^{1/3} \,  \nabla^2_r \ n_G ({\un r},
{\un b}, y = 0) \right) \, = \, 0
\ee
where $n_G$ is the gluon dipole cross section on a single nucleon
obtained from \eq{GM} by expanding it to the lowest non-trivial order
and putting $A=1$. Remembering that
\be
\int_A d^2 b \, = \, A^{2/3} \int_N d^2 b
\ee
we conclude from Eqs. (\ref{sum1}) and (\ref{subtr}) that in the
quasi-classical approximation (see also \cite{KM})
\be\label{sumkt}
\int d^2 k \, \phi_A (y=0, \un k^2) \, = \, A \, 
\int d^2 k \, \phi_N (y=0, \un k^2). 
\ee
Similarly one can show that the Weizs\"{a}cker-Williams gluon
distribution in \eq{wwglue} obeys the same sum rule in the
quasi-classical approximation
\be\label{sumww}
\int d^2 k \, \phi^{WW}_A (y=0, \un k^2) \, = \, A \, 
\int d^2 k \, \phi^{WW}_N (y=0, \un k^2). 
\ee

However, the sum rules of Eqs. (\ref{sumkt}) and (\ref{sumww}) break
down when the non-linear evolution with energy \cite{yuri1,bal} is
included. To see this we first note that for very small $r_T$ one can
use the expression for $N_G$ given by the double logarithmic
approximation \cite{Ryskin,LT,IIM1} (see Sect. IV for details on
similar calculations)
\be\label{Ndla}
 N_G (r_T \approx 0, {\un b}, y) \, = \, \frac{r_T^2 \,
 Q_{s0}^2}{8 \, \sqrt{\pi}} \, \frac{\ln^{1/4} \frac{1}{r_T \,
 Q_{s0}}}{(2 \, \bas \, y)^{3/4}} \, e^{2 \, \sqrt{2 \, \bas \, y \,
 \ln 1/(r_T \, Q_{s0})}}
\ee
with
\be\label{bas}
\bas \, = \, \frac{\as \, N_c}{\pi}.
\ee
Similarly for the proton amplitude $n_G$ we write
\be\label{ndla}
 n_G (r_T \approx 0, {\un b}, y) \, = \, \frac{r_T^2 \,
 \Lambda^2}{8 \, \sqrt{\pi}} \, \frac{\ln^{1/4} \frac{1}{r_T \,
 \Lambda}}{(2 \, \bas \, y)^{3/4}} \, e^{2 \, \sqrt{2 \, \bas \, y \,
 \ln 1/(r_T \, \Lambda)}},
\ee
where now the scale characterizing the proton is given by 
\be\label{psat}
\Lambda^2 \, = \, 4 \, \pi \, \as^2 \frac{1}{S_p} 
\ee
with $S_p$ the cross sectional transverse area of the
proton. Employing the fact that $Q_{s0}^2 = A^{1/3} \Lambda^2$ we can
easily see that the amplitudes in Eqs. (\ref{Ndla}) and (\ref{ndla})
do not satisfy the condition of \eq{subtr} invalidating the sum
rule. In fact using Eqs. (\ref{Ndla}) and (\ref{ndla}) in \eq{subtr}
gives an inequality
\be\label{ineq1}
\lim_{r_T \rightarrow 0} \left( \nabla^2_r \ N_G ({\un r},
{\un b}, y = \ln 1/x) - A^{1/3} \, \nabla^2_r \ n_G ({\un r},
{\un b}, y = \ln 1/x) \right) \, < \, 0.
\ee
\eq{ineq1}, together with a similar equation for $N_G / {\un r}^2$ turn 
the sum rules of Eqs. (\ref{sumkt}) and (\ref{sumww}) into
inequalities
\be\label{inkt}
\int d^2 k \, \phi_A (x, \un k^2) \, \le \, A \, 
\int d^2 k \, \phi_N (x, \un k^2) 
\ee
and
\be\label{inww}
\int d^2 k \, \phi^{WW}_A (x, \un k^2) \, \le \, A \, 
\int d^2 k \, \phi^{WW}_N (x, \un k^2), 
\ee
where the equality is achieved only in the quasi-classical limit. We
conclude that while multiple rescatterings of gluons in
McLerran-Venugopalan model preserve the total number of gluons in a
nuclear wave function at a given rapidity $y$, the quantum evolution
tends reduce the number of gluons in the wave function via gluon
mergers \cite{glr}.

To study nuclear modification of the gluonic wave functions let us
define the unintegrated gluon distributions ratios as
\be\label{rshad}
R_A (x, \un k^2) \, = \, \frac{\phi_A (x,\un k^2)}{A \, \phi_N (x, \un
k^2)} \hspace*{.7cm} \mbox{and} \hspace*{.7cm} R_A^{WW} (x,\un k^2) \,
= \, \frac{\phi^{WW}_A (x, \un k^2)}{A \, \phi^{WW}_N (x, \un k^2)}.
\ee

The sum rules of Eqs. (\ref{sumkt}) and (\ref{sumww}) imply that, in
the quasi-classical approximation, if at some $k_T$ the distribution
function $\phi^{WW}_A (y=0,\un k^2)$ is smaller than $A \, \phi^{WW}_N
(y=0, \un k^2)$, then at some other $k_T$ it should be bigger than $A
\, \phi^{WW}_N (y=0, \un k^2)$. Using the definitions (\ref{rshad}) one 
concludes from \eq{sumww} that if $R_A^{WW} (y=0, \un k^2)$ is below
$1$ at some $k_T$ it is bound to go above $1$ at some other $k_T$ (for
the same value of $x$). From \eq{asym} we can conclude that
\be
R_A (y=0, \un k^2), \, R_A^{WW} (y=0, \un k^2)
\, \rightarrow \, 1, \hspace*{1cm} k_T \,
\rightarrow \, \infty.
\ee
At the same time, when $k_T \ll Q_{s0}$ the saturation effects become
important driving $\phi^{WW}_A (y=0, \un k^2)$ below $A \, \phi^{WW}_N
(y=0, \un k^2)$, or, equivalently, making $R_A^{WW} (y=0, \un k^2) <
1$. Therefore, due to the sum rule of Eqs. (\ref{sumkt}) and
(\ref{sumww}), somewhere in the region of $k_T \gsim Q_{s0}$ the ratio
$R_A^{WW} (y=0, \un k^2)$ should go above one, which corresponds to
enhancement or broadening of the $k_T$ distribution of gluons inside
the nucleus. The same broadening argument applies to $\phi_A (y=0, \un
k^2)$. We have therefore proved that for both gluon distribution
functions calculated in McLerran-Venugopalan model the effects of
saturation and the sum rule (\ref{sumkt}),(\ref{sumww}), while making
$R_A (y=0, \un k^2) < 1$ in the infrared, also require an existence of
a $k_T$-region where $R_A (y=0, \un k^2)$ is above $1$. This
conclusion will be quantified in the next Subsection.

The above argument does not apply to the shadowing ratios $R_A (x, \un
k^2)$ and $R_A^{WW} (x, \un k^2)$ when the effects of quantum
evolution are included. The sum rules (\ref{sumkt}) and (\ref{sumww})
are replaced by inequalities (\ref{inkt}) and (\ref{inww}) which only
require a reduction of the overall number of gluons in the nuclear
wave function at a given rapidity $y$.


\subsection{$k_T$-dependence: Quasi-Classical Approximation}

To investigate the $k_T$-dependence of the unintegrated nuclear gluon
distributions $\phi^{WW}_A (x,\un k^2)$ and $\phi_A (x, \un k^2)$ more
quantitatively and demonstrate the differences of the two distributions
let us study them in McLerran-Venugopalan model
\cite{mv,kjklw}. For that we take the gluon dipole amplitude in the 
Glauber-Mueller approximation \cite{Mueller1} of \eq{GM}. The
high-$k_T$ asymptotic for both $\phi^{WW}_A (x, \un k^2)$ and $\phi_A (x,
\un k^2)$ is given by \eq{asym}.

Inside the saturation region ($k_T \ll Q_{s0}$) one has
\be\label{ktsat}
\phi (x, \un k^2) \, \approx \, 
\frac{2 \, C_F \, S_A}{\as \, (2 \pi)^2} \, 
\frac{k_T^2}{Q_{s0}^2}, \hspace*{1cm} k_T \ll Q_{s0}, 
\ee
and
\be\label{wwsat}
\phi^{WW} (x, \un k^2) \, \approx \, 
\frac{4 \, C_F \, S_A}{\as \, (2 \pi)^2} \, 
\ln \frac{Q_{s0}}{k_T}, \hspace*{1cm} k_T \ll Q_{s0},
\ee
where we assumed for simplicity that the nucleus is cylindrical in
which case its cross sectional area is $S_A = \pi {\cal R}^2$ and
$Q_{s0}$ is given by \eq{sat} with $\rho\, T(\un b)=A/S_A$:
\be\label{satmom}
Q_{s0}^2\,=\,\frac{4\pi\,\as^2\, A}{S_A},\quad \mathrm{cylindrical\;\;
nucleus.}
\ee

In Eqs. (\ref{ktsat}) and (\ref{wwsat}) the difference between the two
gluon distribution functions becomes manifest: $\phi^{WW}_A (x, \un k^2)$
keeps increasing (though only logarithmically) as $k_T$ decreases,
while $\phi_A (x, \un k^2)$ turns over and goes to zero in the
infrared. Still for both distribution functions the ratio $R_A$ goes
to zero as $k_T \rightarrow 0$ since to obtain $R_A$ one has to divide
Eqs. (\ref{ktsat}) and (\ref{wwsat}) by $A \phi_N (x,\un k^2)$ from
\eq{asym}. The sum rules (\ref{sumww}) and (\ref{sumkt}) require a
region of enhancement ($R_A > 1$) at $k_T
\gsim Q_{s0}$. To see that the enhancement really happens one has to 
calculate the next-to-leading twist correction to the high-$k_T$
asymptotic of \eq{asym}. This technique has been applied previously
for quark production in \cite{GJ}. One obtains
\be\label{ktasym}
\phi_A (x, \un k^2) \, = \, \frac{C_F \, S_A \, Q_{s0}^2}{\as \, 
( 2 \pi)^2 \, \un k^2} 
\, \left[ 1 + 2 \, \frac{Q_{s0}^2}{\un k^2} \left( \ln 
\frac{\un k^2}{4 \, \Lambda^2} + 2 \, \gamma -3 \right) + \ldots \right], 
\hspace*{1cm} k_T \rightarrow \infty,
\ee
and
\be\label{wwasym}
\phi^{WW}_A (x, \un k^2) \, = \, \frac{C_F \, S_A \, Q_{s0}^2}{\as \, 
( 2 \pi)^2 \, \un k^2} \, \left[ 1 + \frac{Q_{s0}^2}{\un k^2} \left( \ln 
\frac{k_T}{2 \, \Lambda} + \gamma -1 \right) + \ldots \right], 
\hspace*{1cm} k_T \rightarrow \infty,
\ee
with $\gamma$ the Euler constant. For the ratios $R_A$'s this implies
\be\label{wwra}
R_A \, = \, 1 + 2 \, \frac{Q_{s0}^2}{\un k^2} \left( \ln 
\frac{\un k^2}{4 \, \Lambda^2} + 2 \, \gamma -3 \right) + \ldots
\ee
and
\be\label{ktra}
R_A^{WW} \, = \, 1 + \frac{Q_{s0}^2}{\un k^2} \left( \ln 
\frac{k_T}{2 \, \Lambda} + \gamma -1 \right) + \ldots.
\ee
Therefore the ratios of gluon distributions approach $1$ from above
for both distribution functions at large $k_T$. This of course
indicates the presence of high-$k_T$ enhancement. 

\begin{figure}
\begin{center}
\epsfxsize=10cm
\leavevmode
\hbox{ \epsffile{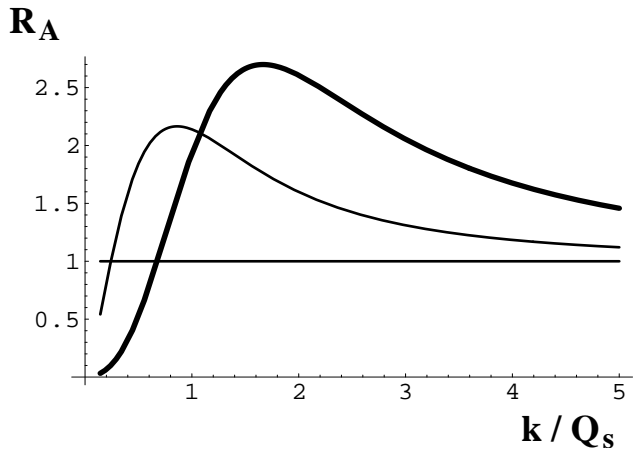}}
\end{center}
\caption{The ratio $R_A$ of unintegrated gluon distributions in the 
nucleus and in the nucleon. The thin line represents the
Weizs\"{a}cker-Williams gluon distribution [\eq{wwglue}] while the
thick line correspond to the more conventional one inspired by
$k_T$-factorization [\eq{ktglue}].}
\label{shad}
\end{figure}

Qualitative plots of ratio $R_A$ for both distribution functions in
McLerran-Venugopalan model are shown in \fig{shad}. The thin line
corresponds to Weizs\"{a}cker-Williams gluon distribution $\phi^{WW}_A
(x, \un k^2)$ while the thick one represents the $k_T$-factorization
distribution $\phi_A (x,\un k^2)$. One can see that in accordance with
Eqs. (\ref{ktsat}) and (\ref{wwsat}) the distribution $\phi_A (x,
\un k^2)$ goes to zero faster than $\phi^{WW}_A (x, \un k^2)$ as $k_T
\rightarrow 0$ in \fig{shad}. In agreement with Eqs. (\ref{ktra}) 
and (\ref{wwra}) $R_A$ for the distribution $\phi_A (x, \un k^2)$ has a
stronger high-$k_T$ enhancement than $R_A^{WW}$ for the distribution
$\phi_A^{WW} (x, \un k^2)$.

Finally, let us point out that the function $R_A$ ($R_A^{WW}$) shown
in \fig{shad} will be modified when quantum evolution is included. Due
to the inequalities of Eqs. (\ref{inkt}) and (\ref{inww}) the total
number of gluons will decrease. As we will see below in Sect. IV the
effects of quantum evolution is to introduce suppression of gluons at
all $k_T$.


\section{Quasi-Classical Approximation: Cronin Effect Only}

\subsection{Gluon Production in pA}

\begin{figure}
\begin{center}
\epsfxsize=10cm
\leavevmode
\hbox{ \epsffile{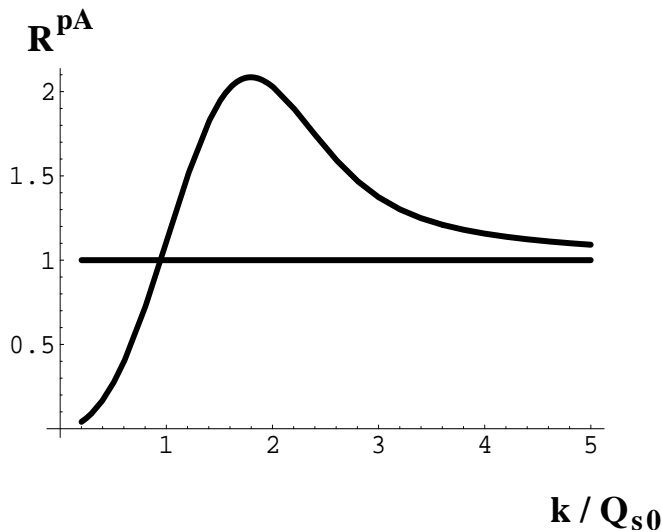}}
\end{center}
\caption{ The ratio $R^{pA}$ for gluons plotted as a function of 
$k_T/Q_{s0}$ in the quasi-classical McLerran-Venugopalan model as
found in \protect\cite{KM}. The cutoff is $\Lambda = 0.2 \, Q_s$.}
\label{cron}
\end{figure}

The problem of gluon production in proton-nucleus collisions in the
quasi-classical approximation (McLerran-Venugopalan model) has been
solved in \cite{KM} (see also \cite{KTS,yuriaa,DM,KW1}). For a
quark-nucleus scattering the production cross section reads \cite{KM}
\begin{eqnarray}\label{pamv}
\frac{d \sigma^{pA}}{d^2 k \ dy} \, = \, \int d^2 b 
\, d^2 x \, d^2 y && \frac{1}{(2 \pi)^2} \, \frac{\as C_F}{\pi^2} \,
\frac{{\underline x} \cdot {\underline y}}{{\underline x}^2 
{\underline y}^2} \, e^{- i {\underline k} \cdot ({\underline x} -
{\underline y})}\nonumber\\
 && \times\left( 1 - e^{- {\un x}^2 Q_{s0}^2\ln(1/x_T\Lambda) /4 } 
- e^{- {\un y}^2  Q_{s0}^2\ln(1/y_T\Lambda) /4 } + e^{- ({\un x} -
{\un y})^2 Q_{s0}^2\ln(1/(\un x-\un y)_T\Lambda) /4 } \right),
\end{eqnarray}
which then has to be convoluted with the light cone wave function of a
quark in a proton. The saturation scale $Q_{s0}^2$ in \eq{pamv} is
given by \eq{sat}. As was shown in \cite{KM} in the approximation when
the logarithmic dependence of exponential factors in \eq{pamv} on the
transverse size is neglected, $\un x^2\ln(1/x_T\Lambda)\approx \un
x^2$, the $x_\perp$ and $y_\perp$ integrations in \eq{pamv} can be
done exactly yielding
\be\label{paqs}
\frac{d \sigma^{pA}}{d^2 k \ dy} \, = \, \frac{\as \, C_F}{\pi^2} \,\int d^2 b 
\, \left\{ - \frac{1}{{\un k}^2} +  \frac{2}{{\un k}^2} \, e^{- {\un k}^2 / 
Q_{s0}^2} + \frac{1}{Q_{s0}^2} \, e^{- {\un k}^2 / Q_{s0}^2} \, \left[
\ln \frac{Q_{s0}^4}{4 \, \Lambda^2 {\un k}^2} + \mbox{Ei} \left( 
\frac{{\un k}^2}{Q_{s0}^2} \right) \right] \right\}, 
\ee
where $\mathrm{Ei}(x)$ is the exponential integral.  Our goal is to
construct the ratio of the number of gluons produced in a pA collision
over the number of gluons produced in a pp collision scaled by the
number of collisions
\be\label{rpa}
R^{pA} (\un k, y) \, = \, \frac{\frac{d \sigma^{pA}}{d^2 k \ dy}}{A \, \frac{d
\sigma^{pp}}{d^2 k \ dy}}.
\ee
In the same approximation in which \eq{paqs} is derived the gluon
production cross section in pp scaled up by $A$ is given by
\be\label{pp}
A \, \frac{d \sigma^{pp}}{d^2 k \ dy} \, = \, \frac{\as \, C_F}{\pi^2} \, 
\int_A d^2 b \, \frac{Q_{s0}^2}{{\un k}^4},
\ee
which can be obtained for instance by taking the $k_T/Q_{s0} \gg 1$
limit of \eq{paqs} and using the fact that $Q_{s0}^2 \sim
A^{1/3}$. For a cylindrical nucleus the impact parameter ${\un b}$
integration would just give a factor of $S_A$. Using Eqs. (\ref{paqs})
and (\ref{pp}) in \eq{rpa} we then obtain
\be\label{rpa1}
R^{pA} (k_T) \, = \, \frac{{\un k}^4}{Q_{s0}^2} \, 
\left\{ - \frac{1}{{\un k}^2} +  \frac{2}{{\un k}^2} \, e^{- {\un k}^2 / 
Q_{s0}^2} + \frac{1}{Q_{s0}^2} \, e^{- {\un k}^2 / Q_{s0}^2} \, \left[
\ln \frac{Q_{s0}^4}{4 \, \Lambda^2 {\un k}^2} + \mbox{Ei} \left( 
\frac{{\un k}^2}{Q_{s0}^2} \right) \right] \right\}.
\ee
The ratio $R^{pA} (k_T)$ is plotted in \fig{cron} for $\Lambda = 0.2 \
Q_{s0}$. It clearly exhibits an enhancement at high-$k_T$ typical of
Cronin effect \cite{Cronin}. Similar conclusions regarding formula
(\ref{pamv}) have been reached earlier in \cite{KNST}.

It is worth noting that expanding $R^{pA} (k)$ from \eq{rpa1} in the
powers of $Q_{s0} / k_T$ (``twists'') yields a series with only
positive terms
\be\label{borel}
R^{pA} (k_T) \, = \, 1 + 2 \, \frac{Q_{s0}^2}{\un k^2} + 6
\,\frac{Q_{s0}^4}{\un k^4} + 24 \, \frac{Q_{s0}^6}{\un k^6} + \ldots \, = \, 
\sum_{n=0}^\infty \, n! \, \left( \frac{Q_{s0}^2}{\un k^2} \right)^n. 
\ee
The series (\ref{borel}) is divergent, but it is Borel resummable with
the sum given by \eq{rpa1}, though not all terms in \eq{rpa1} can be
reconstructed by Borel resummation procedure.

To establish whether inclusion of the correct transverse size
dependence in the exponents of \eq{pamv} would change the conclusion
about Cronin effect let us study the high-$k_T$ asymptotic of
\eq{pamv}. A simple calculation yields
\ben
\frac{d \sigma^{pA}}{d^2 k \ dy} \, = \, \frac{\as \, C_F}{\pi^2} \,
\int d^2 b \, \frac{Q_{s0}^2}{{\un k}^4} \, \left[ 
\left( \ln \frac{{\un k}^2}{4 \, \Lambda^2} + 2 \, \gamma - 1 \right) + \right.
\een
\be\label{paas1} 
+ \left.  \frac{Q_{s0}^2}{4 \, {\un k}^2} \, \left( 6 \, \ln^2 
\frac{{\un k}^2}{4 \, \Lambda^2} - 8 \, (4 - 3 \gamma) \,
\ln \frac{{\un k}^2}{4 \, \Lambda^2} + 29 + 24 \, \gamma^2 - 64 \, \gamma \right) 
+ \ldots \right], \hspace*{1cm} k_T \rightarrow \infty.
\ee
For a cylindrical nucleus, keeping only the leading logarithmic ($\ln
\frac{{\un k}^2}{\Lambda^2}$) terms in the parentheses of
\eq{paas1} we obtain
\be\label{qclt}
R^{pA} (k_T) \, = \, 1 + \frac{3}{2} \, \frac{Q_{s0}^2}{{\un k}^2} \,
\ln
\frac{{\un k}^2}{\Lambda^2} + \ldots, \hspace*{1cm} k_T \rightarrow \infty
\ee
indicating that $R^{pA}$ approaches $1$ from above at high $k_T$, which
is typical of Cronin enhancement. We therefore conclude that in the
framework of the quasi-classical approximation employed in \cite{KM}
the ratio $R^{pA}$ is less than $1$ at small $k_T \lsim Q_{s0}$ and
has Cronin enhancement at high $k_T \gsim Q_{s0}$.

As can be seen from Eqs. (\ref{rpa1}) and (\ref{qclt}), the position
of Cronin maximum is determined by the saturation scale, such that
$k_{\rm max} = \beta \, Q_{s0}$, where $\beta$ is some weakly
increasing function of $\ln Q_{s0}/\Lambda$. The height of the maximum
is given by $R^{pA} (k_{\rm max}) = R^{pA} (\beta \, Q_{s0})$.
Substituting $k_T = \beta \, Q_{s0}$ in \eq{rpa1} we observe that the
height of Cronin maximum scales like
\be\label{max}
R^{pA} (\beta \, Q_{s0}) \, \sim \, \ln \frac{Q_{s0}}{\Lambda} +
\mbox{const} \, \sim \, \ln A + \mbox{const}'.
\ee
Since, for realistic off-central collisions $A$ is replaced by the
number of participants $N_{part}$, we conclude from \eq{max} that in
the quasi-classical approximation considered here the $k_T$-position
and the height of the Cronin peak should increase with centrality of
the $pA$ collision.


\subsection{$k_T$-Factorization}

Let us now show that it is possible to rewrite \eq{pamv} in a
$k_T$-factorized form \cite{glr,Braun2,KT}. Repeating the steps
outlined in Sect. IV of \cite{KT} we first perform one of the
transverse coordinate integrations in \eq{pamv} rewriting it as
\be\label{pamv2}
\frac{d \sigma^{pA}}{d^2 k \ dy} \, = \, \frac{1}{2 \pi^2} \, 
\frac{\as C_F}{\pi} \, \int d^2 b \, d^2 z \, 
e^{- i {\un k} \cdot {\un z}} \, \left[ 2 \, i \, 
\frac{{\un z} \cdot {\un k}}{{\un z}^2 {\un k}^2} -  
\ln \frac{1}{z_T \Lambda} \right] \, N_G ({\un z}, {\un b}, 0),
\ee
where $N_G ({\un z}, {\un b}, 0)$ is given by \eq{GM}. Using the
fact that $N_G (z=0, {\un b}, 0) = 0$ we write \eq{pamv2} as
\be\label{pamv3}
\frac{d \sigma^{pA}}{d^2 k \ dy} \, = \, \frac{1}{2 \pi^2} \, 
\frac{\as C_F}{\pi} \, \frac{1}{{\un k}^2} \, \int d^2 b \, d^2 z \, 
N_G ({\un z}, {\un b}, 0) \, \nabla^2_z \, \left( e^{- i {\un k}
\cdot {\un z}} \, \ln \frac{1}{z_T \Lambda} \right).
\ee
Let us denote the forward scattering amplitude of a gluon dipole of
transverse size ${\un r}$ on a single nucleon (proton) integrated over
the impact parameter ${\un b}'$ of the dipole measured with respect to
the proton by
\be\label{dipp}
\int d^2 b' \, n_G ({\un r}, {\un b}', y=0) \, = \, \pi \, \as^2 \,
{\un r}^2 \, \ln \frac{1}{ r_T \, \Lambda}.
\ee
\eq{dipp} is obtained by expanding \eq{GM} at the leading order and taking
$A=1$. It corresponds to the two gluon exchange interaction between
the dipole and the proton. In the quasi-classical Glauber-Mueller
approximation in which \eq{GM} is derived each nucleon exchanges only
two gluons with the dipole \cite{Mueller1,kjklw}. Therefore \eq{dipp}
is a natural reduction of \eq{GM} to a single nucleon case.

\begin{figure}
\begin{center}
\epsfxsize=5cm
\leavevmode
\hbox{\epsffile{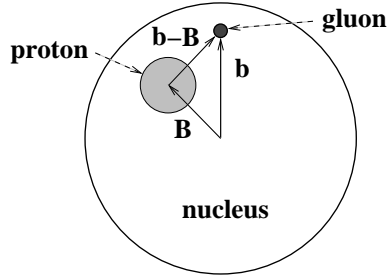}}
\end{center}
\caption{Gluon production in pA collisions as seen in the transverse 
plane. To make the picture easier to read the gluon is placed far away
from the proton which is highly unlikely to happen in real life.}
\label{pa}
\end{figure}

With the help of \eq{dipp} we rewrite \eq{pamv3} as \cite{KT}
\be\label{pamvkt}
\frac{d \sigma^{pA}}{d^2 k \ dy} \, = \, \frac{C_F}{\as \, 
\pi \, (2 \pi)^3} \, 
\frac{1}{{\un k}^2} \, \int d^2 B \, d^2 b \, d^2 z \, 
\nabla^2_z \, n_G ({\un z}, {\un b} - {\un B}, 0) \, e^{- i {\un k}
\cdot {\un z}} \, \nabla^2_z \, N_G ({\un z}, {\un b}, 0).
\ee
Now ${\un B}$ is the impact parameter of the proton with respect to
the center of the nucleus and ${\un b}$ is the impact parameter of the
gluon with respect to the center of the nucleus as shown in \fig{pa}.
\eq{pamvkt} is the expression for gluon production one would write in 
the $k_T$-factorization approach \cite{Braun2}. To see this explicitly
let us rewrite \eq{pamvkt} in terms of the unintegrated gluon
distribution function from \eq{ktglue}. One easily derives
\be\label{kt}
\frac{d \sigma^{pA}}{d^2 k \ dy} \, = \, \frac{2 \, \as}{C_F} \, 
\frac{1}{{\un k}^2} \, \int d^2 q \, \phi_p ({\un q}) \, \phi_A ({\un k} 
- {\un q}), 
\ee
which is the same formula as obtained in $k_T$-factorization approach
\cite{glr,Ryskin,Braun2}. $\phi_p$ is defined as unintegrated gluon 
distribution of the proton given by \eq{ktglue} with $n_G$ instead
of $N_G$ on the right hand side.  \eq{kt} demonstrates that the
gluon production cross section in pA can be expressed in terms of the
gluon distribution (\ref{ktglue}) in a rather straightforward way
\cite{KT}. Somehow it is the distribution (\ref{ktglue}) and not the
Weizs\"{a}cker-Williams distribution (\ref{wwglue}) that enters
\eq{kt}.

\eq{kt} demonstrates that, at least in the framework of 
McLerran-Venugopalan model, the multiple rescattering leading to
Cronin enhancement in pA can be incorporated in the gluon distribution
functions \cite{KM,yuriaa}. There is no clear distinction between the
nuclear wave function effects and the Glauber-type rescatterings in
the nucleus. Anti-shadowing present in the gluon distribution function
$\phi_A ({\un k})$ as shown in \fig{shad} simply translates into
Cronin effect of \fig{cron} via \eq{kt}.

In the quasi-classical approximation of McLerran-Venugopalan model one
can prove a sum rule for the gluon production cross section in $pA$
similar to the sum rule we proved for gluon distributions in
Sect.~IIB. To prove the sum rule we note that \eq{pamvkt} implies that
\be
\int d^2 k \, {\un k}^2 \, \frac{d \sigma^{pA}}{d^2 k \ dy} \, = \, 
\frac{C_F}{\as \, 2 \pi^2} \, 
\, \int d^2 B \, d^2 b \, 
\left[ \nabla^2_z \, n_G ({\un z}, {\un b} - {\un B}, 0) \right] 
\bigg|_{{\un z}=0}\, 
\left[ \nabla^2_z \, N_G ({\un z}, {\un b}, 0) \right] 
\bigg|_{{\un z}=0}.
\ee
For Glauber-Mueller $N_G$ from \eq{GM} and for $n_G$ from \eq{dipp}
the following condition is satisfied
\be\label{subtr'}
\lim_{z_T \rightarrow 0} \left\{ \left[ \nabla^2_z \, 
n_G ({\un z}, {\un b} - {\un B}, 0) \right] \,
\left[ \nabla^2_z \, N_G ({\un z}, {\un b}, 0) \right] - A^{1/3}
\left[ \nabla^2_z \, 
n_G ({\un z}, {\un b} - {\un B}, 0) \right] \, 
\left[ \nabla^2_z \, n_G ({\un z}, {\un b}, 0) \right] \right\} \, = \, 0.
\ee
The impact parameter integration in pA will give an extra factor of
$A^{2/3}$ as compared to pp. Together with \eq{subtr'} this gives
\be\label{sumr}
\int d^2 k \, {\un k}^2 \, \frac{d \sigma^{pA}_{MV}}{d^2 k \ dy} \, = \, 
A \, \int d^2 k \, {\un k}^2 \, \frac{d \sigma^{pp}_{MV}}{d^2 k \ dy}
\ee
in the quasi-classical approximation. 

Similar to the sum rule proved in Sect. II for gluon distribution
functions, the sum rule (\ref{sumr}) insures that if the
quasi-classical gluon production cross section in pA collisions is, in
some region of $k_T$, smaller than $A$ times the gluon production
cross section in pp than there should be some other region of $k_T$ in
which their roles are reversed. For $R^{pA}$ defined in \eq{rpa} that
means that if, in some region of $k_T$, it is less than $1$ there must
be some other region of $k_T$ in which it is greater than $1$. Of
course the ${\un k}^2$ factors in \eq{sumr} make the quantitative
amounts of suppression and enhancement very different from the ones
dictated by, for instance, the sum rule of Sect. II.

In the quasi-classical approximation for the gluon production in pA
considered above $R^{pA}$ is below $1$ at $k_T \lsim
Q_{s0}$. Expanding \eq{rpa1} for $k_T \ll Q_{s0}$ we write
\be\label{rpa2}
R^{pA} (k) \, \approx \, \frac{{\un k}^2}{Q_{s0}^2} \, \ll \, 1
\hspace*{1cm} \mbox{if} \hspace*{1cm} k_T \ll Q_{s0}.
\ee
\eq{rpa2}, together with the sum rule (\ref{sumr}) imply that there must 
exist a region of $k_T$ with a Cronin-like enhancement of gluon
production, which is demonstrated by the full answer plotted in
\fig{cron}.


\section{Including Small-$x$ Evolution: Suppression at all $p_T$}

\subsection{Including Small-$x$ Evolution}

As the energy of the collisions increases quantum evolution
corrections become important. For produced particles with the same
$k_T$ higher energy implies smaller effective Bjorken $x$ meaning that
the quantum corrections of the type $\as \ln 1/x$ should be
resummed. These corrections can be resummed by the BFKL equation
\cite{BFKL}, which calculates the contribution of the hard (perturbative) 
pomeron. However, as energy increases multiple pomeron exchanges
become important, resulting in a more complicated small-$x$ evolution
\cite{JKLW,FILM}. In \cite{yuri1,bal} an equation was constructed
which resums multiple pomeron exchanges for a forward amplitude of a
$q\bar q$ dipole scattering on a nucleus in the large $N_c$ limit. The
forward amplitude $N ({\un r}, {\un b}, Y)$ of a dipole of
transverse size ${\un r}$ scattering at impact parameter ${\un b}$ and
rapidity $Y$ was normalized such that the total $q{\bar q} A$ cross
section was given by
\be\label{norm}
\sigma^{q{\bar q} A}_{tot} \, = \, 2 \int d^2 b \, 
 N ({\underline r},{\underline b} , Y).
\ee
The evolution equation for $N(\un r,\un b ,Y)$
 closes only in the large-$N_c$ limit
of QCD \cite{bal,JKLW} and reads \cite{yuri1,yuri2,dip}
\ben
  N({\underline x}_{01},{\underline b}, Y) = N ({\underline
  x}_{01},{\underline b}, Y=0) \, e^{- \frac{4 \alpha
  C_F}{\pi} \ln \left( \frac{x_{01}}{\rho} \right) Y} +
  \frac{\alpha C_F}{\pi^2} \int_0^Y d y  \, e^{ - \frac{4
  \alpha C_F}{\pi} \ln \left( \frac{x_{01}}{\rho} \right) (Y - y)}
\een
\be\label{eqN}
\times \int_\rho d^2 x_2 \frac{x_{01}^2}{x_{02}^2 x_{12}^2} \, [ 2
  \, N ({\underline x}_{02},{\underline b} + \frac{1}{2} {\underline
  x}_{12}, y) - N ({\underline x}_{02},{\underline b} + \frac{1}{2}
  {\underline x}_{12}, y) \, N ({\underline x}_{12},{\underline b} +
  \frac{1}{2} {\underline x}_{02}, y) ] ,
\ee
with the initial condition given by $N ({\underline
x}_{01},{\underline b}, Y=0)$ taken to be of Mueller-Glauber form
\cite{Mueller1} in \cite{yuri1}:
\begin{equation}\label{gla}
   N ({\underline x}_{01},{\underline b}_0, Y=0) = 1 - e^{ -
   {\underline x}_{01}^2 Q_{0s}^{\mathrm{quark}\, 2}\ln(1/x_{01T}\Lambda) / 4},
\end{equation}
where 
\be\label{xqs}
N_c\,Q_{0s}^{\mathrm{quark}\, 2} \, = \,  C_F\, Q_{s0}^2.
\ee

In \cite{KT} it was shown how to resum the effects of nonlinear
evolution of \eq{eqN} for gluon production in DIS. In the
quasi-classical approximation the gluon production in DIS is given by
a formula similar to \eq{pamv} \cite{yuridiff}. That formula can also
be recast in a $k_T$ factorized from of \eq{pamvkt} \cite{KT}. As was
proven in \cite{KT} in order to include quantum evolution (\ref{eqN})
in \eq{pamvkt} for DIS one has to make replacements. First, one has to
replace $N_G ({\un z}, {\un b}, 0)$ in \eq{pamvkt} by the forward
quark dipole amplitude using the following expression valid in the
large-$N_c$ limit \cite{KT}
\be\label{nga}
N_G ({\un z}, {\un b}, y) = 2 N ({\un z}, {\un
b}, y) - N ({\un z}, {\un b}, y)^2, 
\ee
where $N ({\un z}, {\un b}, y)$ is the forward scattering amplitude
of a $q\bar q$ dipole on a nucleus evolved by nonlinear equation
(\ref{eqN}). Then one has to replace $n_G ({\un z}, {\un b}, 0)$ by
$n_G ({\un z}, {\un b}, Y-y)$ evolved just by the linear part of
\eq{eqN} (the BFKL equation \cite{BFKL}). Here $Y$ is the total
rapidity interval between the projectile (virtual photon) and target
nucleus in a DIS collision. The initial conditions for both $N_G$
and $n_G$ evolution are given by $N_G ({\un z}, {\un b}, 0)$ and
$n_G ({\un z}, {\un b}, 0)$ correspondingly. 

Since both pA and DIS are considered here as scatterings of an
unsaturated projectile (proton or $q\bar q$ pair) on a saturated
target (nucleus) with the gluon production in the quasi-classical
limit given by the same \eq{pamvkt}, we may conjecture that inclusion
of quantum evolution (\ref{eqN}) in gluon production cross section is
done similarly for both processes. We therefore write
\be\label{paev}
\frac{d \sigma^{pA}}{d^2 k \ dy} \, = \, \frac{C_F}{\as \, 
\pi \, (2 \pi)^3} \, 
\frac{1}{{\un k}^2} \, \int d^2 B \, d^2 b \, d^2 z \, 
\nabla^2_z \, n_G ({\un z}, {\un b} - {\un B}, Y-y) \, e^{- i {\un k}
\cdot {\un z}} \, \nabla^2_z \, N_G ({\un z}, {\un b}, y),
\ee
where $Y$ is the total rapidity interval between the proton and the
nucleus. Just like in DIS $N_G$ in \eq{paev} is given by \eq{nga},
where $N$ should be found from \eq{eqN}, while $n_G$ should be
determined from the linear part of \eq{paev} (BFKL) with the initial
conditions given by \eq{dipp}. \eq{paev} is exact if the proton is
modeled as a diquark--quark pair \cite{diq}, in which case it would be
identical to $q\bar q$ pair produced by a virtual photon in DIS. In
general case \eq{paev} remains a well-motivated conjecture.

Like in Sect. II the sum rule (\ref{sumr}) breaks down once non-linear
evolution \cite{yuri1,bal} is included in the way shown in
\eq{paev}. Using the double logarithmic expressions (\ref{Ndla}) and
(\ref{ndla}) modifies \eq{subtr'} into
\begin{eqnarray}\label{ineq'}
\lim_{z_T \rightarrow 0}  \left\{ \left[ \nabla^2_z \, 
n_G ({\un z}, {\un b} - {\un B}, Y-y) \right] \, 
\right. && \left[ \nabla^2_z \, N_G ({\un z}, {\un b}, y) \right] 
\nonumber\\
&& - A^{1/3} \left. \left[ \nabla^2_z \, 
n_G ({\un z}, {\un b} - {\un B}, Y-y) \right] \, 
\left[ \nabla^2_z \, n_G ({\un z}, {\un b}, y) \right] \right\} \, < \, 0
\end{eqnarray}
turning the sum rule of \eq{sumr} into an inequality for the cross
section from \eq{paev}
\be\label{inequa}
\int d^2 k \, {\un k}^2 \, \frac{d \sigma^{pA}}{d^2 k \ dy} \, \le \, 
A \, \int d^2 k \, {\un k}^2 \, \frac{d \sigma^{pp}}{d^2 k \ dy}.
\ee
Again the effect of quantum evolution is to reduce the total number of
gluons at a given rapidity, though now it is shown for the case of
gluon production weighted by $k_T^2$. Let us now study in detail how
this suppression sets in for various regions of $k_T$.

In the following we are going to study effects of evolution equation
(\ref{eqN}) on the gluon spectrum and on $R^{pA}$. Our goal is to
determine whether $R^{pA}$ preserves the shape shown in
\fig{cron} with Cronin maximum and low-$k_T$ suppression, or quantum
evolution would modify this shape introducing extra suppression. Below
we will first study the effects of quantum evolution at high-$k_T$,
$k_T \gsim Q_s$, showing that evolution introduces suppression
($R^{pA} < 1$) in that region. We will then proceed by studying the
fate of the Cronin peak ($k_T \sim Q_s$) as evolution sets in. We will
show that the Cronin maximum will decrease with the onset of evolution
and would eventually disappear. We will then argue that suppression
persists for $k_T \ll Q_s$ when evolution effects are included. We
will end the section by constructing a toy model summarizing our
conclusions. 

To simplify the discussion we will consider a cylindrical nucleus for
which \eq{paev} reduces to
\be\label{paevc}
\frac{d \sigma^{pA}}{d^2 k \ dy} \, = \, \frac{C_F}{\as \, 
\pi \, (2 \pi)^3} \, 
\frac{S_p \, S_A}{{\un k}^2} \, \int d^2 z \, 
\nabla^2_z \, n_G ({\un z}, Y-y) \, e^{- i {\un k}
\cdot {\un z}} \, \nabla^2_z \, N_G ({\un z}, y),
\ee
with $S_p$ the cross sectional area of the proton.


\subsection{Leading Twist Effects}

\subsubsection{Leading Twist Gluon Production Cross Section}

\begin{figure}
\begin{center}
\epsfxsize=7cm
\leavevmode
\hbox{\epsffile{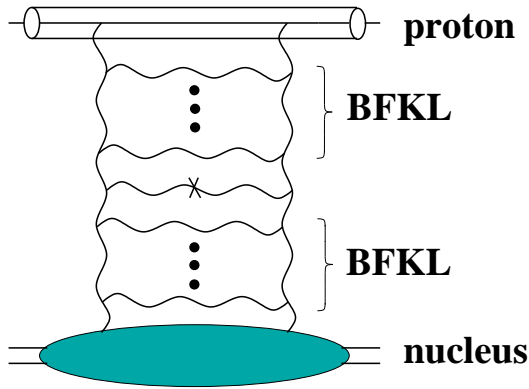}}
\end{center}
\caption{Gluon production in pA collisions at the leading twist level 
(see text).}
\label{incllt}
\end{figure}

We start by exploring the leading high-$k_T$ behavior of the gluon
spectrum given by \eq{paevc}. At very high $k_T$ the integral in
\eq{paevc} is dominated by small values of $z_T$. Therefore we can 
neglect the quadratic term in the evolution equation for $N_G$
(\ref{eqN}) leaving only the linear part -- the BFKL evolution with
initial conditions for a gluon dipole given by \eq{GM}. The
corresponding Feynman diagram is shown in \fig{incllt}. The solution
of the BFKL equation is well-known and reads
\be\label{ltNG}
N_{G \, 1} ({\un z}, y) \, = \, \int \frac{d \lambda}{2 \pi i} \,
C_\lambda^A
\, (z_T \, Q_{s0})^\lambda \, e^{2 \, \bas \, \chi (\lambda) \, y}
\ee
with
\be\label{chi}
\chi (\lambda) \, = \, \psi (1) - \frac{1}{2} \, \psi \left( 1 - 
\frac{\lambda}{2}\right) - \frac{1}{2} \, \psi \left( 
\frac{\lambda}{2}\right),
\ee
$\bas$ defined in \eq{bas} and $Q_{s0}$ for a cylindrical nucleus
given by \eq{satmom}.
The coefficient $C_\lambda^A$ is fixed from the initial conditions at
$y=0$ given by \eq{GM}. Then for small $z_T < 1/Q_{s0}$
\be\label{ca}
C_\lambda^A \, & = & \, \sum_{n=1}^\infty \, \sum_{m=0}^n \, 
\frac{(-1)^{n+1}}{4^n \, (n-m)! \, (2 n - \lambda)^{m+1}} \, 
\ln^{n-m} \frac{Q_{s0}}{\Lambda} \nonumber \\
& = & \,\sum_{n=1}^\infty \,
\frac{1}{4^n \, n! \, (\lambda - 2 n)^{n+1}} \, \left( 
\frac{Q_{s0}}{\Lambda} \right)^{2 n - \lambda} \, 
\Gamma\left( 1+n , \, (2 n - \lambda) \, \ln \frac{Q_{s0}}{\Lambda} \right).
\ee
Similarly for the gluon dipole cross section on the proton we write
\be\label{ltnG}
n_G ({\un z}, y) \, = \, \int \frac{d \lambda}{2 \pi i} \, C_\lambda^p
\, (z_T \, \Lambda)^\lambda \, e^{2 \, \bas \, \chi (\lambda) \, y},
\ee
where the scale characterizing the proton $\Lambda$ is given by
\eq{psat}. The coefficient $C_\lambda^p$ is obtained by requiring that 
\eq{ltnG} reduces to \eq{dipp} when $y=0$. For $z_T < 1/\Lambda$ we derive
\be\label{cp}
C_\lambda^p \, = \, \frac{1}{4 \, (\lambda -2)^2}.
\ee
(We have identified the non-perturbative scale characterizing the
proton (\ref{psat}) with the infrared cutoff employed earlier in
\eq{dipp}.)
Substituting Eqs. (\ref{ltNG}) and (\ref{ltnG}) into
\eq{paevc} and integrating over $\un z$ yields
\cite{DPT,ELR}
\ben
\frac{d \sigma^{pA}}{d^2 k \ dy} \bigg|_{LT} \, = \, 
\frac{C_F \, S_p \, S_A}{4 \, \as \, (2 \pi)^3} \, 
\int \frac{d \lambda}{2 \pi i} \, \frac{d \lambda'}{2 \pi i} \,
\lambda^2 \, \lambda'^2 \, C_\lambda^A \, C_{\lambda'}^p \, 
2^{\lambda + \lambda'} \, \frac{\Gamma \left( -1 + \frac{\lambda +
\lambda'}{2} \right)}{\Gamma \left( 2 - \frac{\lambda +
\lambda'}{2} \right)} \, \left( \frac{Q_{s0}}{k_T} \right)^\lambda \,  
\left( \frac{\Lambda}{k_T} \right)^{\lambda'} \, 
\een
\be\label{lt}
\times \, e^{2 \, \bas \, [\chi (\lambda) \, 
y + \chi (\lambda') \, (Y-y) ]}.
\ee
\eq{lt} gives the leading twist expression for the gluon production 
cross section in pA collisions and is illustrated in \fig{incllt}. 

The difference between \eq{lt} and Eq. (13) of
\cite{DPT} is in gamma-functions in the integrand. The difference
manifests itself at the order of higher twists, where the gluon
distributions used in \cite{DPT}, if taken at $y=0$ and used in
inverted \eq{ktglue} to obtain $n_p$, would yield higher twist
corrections (higher powers of $r_T$) to the right hand side of
\eq{dipp}, which should not be there in the two-gluon exchange
approximation corresponding to $y=0$ limit.


\subsubsection{Double Logarithmic Approximation: Monojet Versus Dijet 
and First Signs of High-$p_T$ Suppression}

To derive the high-$k_T$ behavior of the gluon production cross
section in \eq{lt} we have to evaluate the integrals in it by saddle
point method. When $k_T \gg Q_{s0} , \, \Lambda$ we approximately
write
\be\label{spd}
\chi (\lambda) \, \approx \, \frac{1}{2 - \lambda}.
\ee
Then the saddle points are given by
\be\label{spd1}
\lambda_{sp} \, = \, 2 - \sqrt{\frac{2 \, \bas \, y}{\ln (k_T/Q_{s0}) } }
\ee
and
\be\label{spd2}
\lambda_{sp}' \, = \, 2 - \sqrt{\frac{2 \, \bas \, (Y-y)}{\ln (k_T/\Lambda)}}.
\ee
Integrating over $\lambda$ and $\lambda'$ around the saddle points
(\ref{spd1}) and (\ref{spd2}) in \eq{lt} yields gluon production cross
section in double logarithmic approximation (DLA) \cite{Ryskin}
\ben
\frac{d \sigma^{pA}}{d^2 k \ dy} \bigg|_{DLA} \, \approx \, 
\frac{C_F \, S_p \, S_A}{\as \, (2 \pi)^4} \, 
\frac{Q_{s0}^{2} \, \Lambda^2}{{\un k}^4} \, \frac{1}{2 \, \bas} \, 
\left[ \frac{\ln \frac{k_T}{Q_{s0}} \, \ln \frac{k_T}{\Lambda}}{y^3 \, 
(Y-y)^3} 
\right]^{1/4} \, \left( \sqrt{\frac{y}{\ln \frac{k_T}{Q_{s0}}}} + 
\sqrt{\frac{Y-y}{ \ln \frac{k_T}{\Lambda}}} \right)
\een
\be\label{dla}
\times \exp\left( 2 \sqrt{2 \, \bas \, y \, \ln \frac{k_T}{Q_{s0}}} +
 2 \sqrt{2 \, \bas \, (Y-y) \, \ln \frac{k_T}{\Lambda}} \ \right).
\ee
To understand \eq{dla} let us first construct gluon distribution
function 
\be\label{collg}
x G_A (x, Q^2) \, = \, \int^{Q^2}_{\Lambda^2} d k^2_T \, \phi^A (x,\un k^2)
\ee
in the same double logarithmic approximation \cite{mq}. Using
\eq{ltNG} in Eqs. (\ref{ktglue}) and (\ref{collg}) we obtain in the
double logarithmic approximation
\be\label{dlag}
x G_A (x, Q^2) \, = \, \frac{C_F \, S_A \, Q_{s0}^2}{\as \, (2 \pi)^3} \,
2 \sqrt{\pi} \, \frac{\ln^{1/4} \frac{Q}{Q_{s0}}}{(2 \, \bas \, 
\ln (1/x))^{3/4}} \, e^{2 \sqrt{2 \, \bas \, \ln (1/x) \, \ln (Q/Q_{s0})}}.
\ee
Since the above gluon distribution is obtained in the DLA (large
$Q^2$) limit of the BFKL equation, it can also be obtained by taking
the small-$x$ limit of the DGLAP equation \cite{DGLAP}. One can
explicitly check that with the help of \eq{dlag} and an analogous one
for the proton gluon distribution $xG_p$, \eq{dla} can be rewritten as
\ben
\frac{d \sigma^{pA}}{d^2 k \ dy} \, = \, 
\frac{2 \, \as}{C_F \, {\un k}^2} \, \left[ 
xG_p (x=e^{-Y+y}, {\un k}^2) \,
\frac{\partial}{\partial k_T^2} \, x G_A (x=e^{-y}, {\un k}^2) \, + \right.
\een
\be\label{coll}
\left. + \, xG_A (x=e^{-y}, {\un k}^2) \,
\frac{\partial}{\partial k_T^2} \, x G_p (x=e^{-Y+y}, {\un k}^2) \right].
\ee
\eq{coll} can be obtained directly by using \eq{dlag} in \eq{kt} 
and assuming that the $q$-integration in \eq{kt} is dominated by the
regions near ${\un q} = 0$ and ${\un q} = {\un k}$
\cite{glr,Ryskin}. As was shown in detail in \cite{BM}, \eq{coll} can be 
reduced to
\ben
\frac{d \sigma^{pA}}{d^2 k \ dy} \, = \, \frac{2}{\pi \, N_c \, C_F} \, 
\left( \int_{e^{-Y+y}}^1 
\frac{d x_1}{x_1} \, x_1 G_p (x_1, {\un k}^2) \, x G_A (x=e^{-y}, {\un k}^2)
\, + \right.
\een
\be\label{coll2}
+ \left. \int_{e^{-y}}^1 \frac{d x_1}{x_1} \, x G_p (x=e^{-Y+y}, {\un
k}^2) \, x_1 G_A (x_1, {\un k}^2) \right) \,
\frac{d {\hat \sigma}^{gg\rightarrow gg}}{d^2 k},
\ee
which is the standard dijet production cross section derived in
collinear factorization approximation (see e.g. \cite{BM}). (Of course
one of the jets' momentum in \eq{coll2} is integrated over.) 
Therefore, we have started with a single jet production cross section
given by $k_T$-factorized expression (\ref{lt}) with BFKL gluon
distributions and demonstrated that in the large $k_T$ limit it
reduces to the conventional dijet production cross section
(\ref{coll2}) given by collinear factorization with DGLAP-evolved
structure functions.\footnote{We thank Al Mueller for encouraging one
of the authors (Yu. K.) to verify this correspondence explicitly
several years ago.}

Before we continue let us study $R^{pA}$ given by the cross section of
\eq{dla}. The naive expectation for the high-$k_T$ limit of the 
leading twist gluon production cross section would be that $R^{pA} =
1$. However, already at the level of approximation employed in
\eq{dla} this is not quite the case. To see this let us first write down 
an expression for the gluon production cross section in pp collisions
in the leading twist DLA approximation. It is obtained by replacing
$Q_{s0}$ and $S_A$ in \eq{dla} by $\Lambda$ and $S_p$
correspondingly. We obtain
\be\label{dlapp}
\frac{d \sigma^{pp}}{d^2 k \ dy} \bigg|_{DLA} \, \approx \, 
\frac{C_F \, S_p^2}{\as \, (2 \pi)^4} \
\frac{\Lambda^4}{{\un k}^4} \, \frac{1}{2 \, \bas} \, 
\frac{\sqrt{y} + \sqrt{Y-y}}{y^{3/4} \, (Y-y)^{3/4}} 
\, \exp\left[ 2 \sqrt{2 \, \bas \, \ln \frac{k_T}{\Lambda}} \left( \sqrt{y} 
+ \sqrt{Y-y} \right) \right].
\ee
To calculate $R^{pA}$ we note that since $S_A = A^{2/3} S_p$ one
concludes from Eqs. (\ref{satmom}) and (\ref{psat}) that $Q_{s0}^2 
= A^{1/3} \Lambda^2$. Using Eqs. (\ref{dla}) and (\ref{dlapp}) in
\eq{rpa} yields
\ben
R^{pA} (k_T, y) \bigg|_{k_T \gg Q_s} \, = \, \frac{\left( \ln
\frac{k_T}{Q_{s0}} \, \ln
\frac{k_T}{\Lambda} \right)^{1/4}}{\sqrt{y} + \sqrt{Y-y}} \, 
\left( \sqrt{\frac{y}{\ln \frac{k_T}{Q_{s0}}}} + 
\sqrt{\frac{Y-y}{ \ln \frac{k_T}{\Lambda}}} \right) 
\een
\be\label{dlar}
\times \, \exp\left[ 2 \sqrt{2 \, \bas \, y} \left( \sqrt{\ln
\frac{k_T}{Q_{s0}}} - \sqrt{\ln \frac{k_T}{\Lambda}} \right) \right],
\ee
where $Q_{s} = Q_{s}(y)$ is the full energy dependent saturation
scale, which reduces to $Q_{s0}$ at $y=0$.  Defining
\be\label{xidef}
\xi \, \equiv \, \left( \frac{\ln \frac{k_T}{Q_{s0}}}{\ln \frac{k_T}{\Lambda}}
\right)^{1/4}
\ee
we rewrite \eq{dlar} as
\be\label{rxi}
R^{pA} (\xi, y) \bigg|_{\xi < 1} \, = \, \frac{\frac{1}{\xi}
\, \sqrt{y} + \xi \, \sqrt{Y-y}}{\sqrt{y} + \sqrt{Y-y}} \, \exp \left[ 
 - \, 2 \, \sqrt{2 \, \bas \, y  \, 
\frac{1-\xi^2}{1+\xi^2} \, \ln \frac{Q_{s0}}{\Lambda} } \, \right]
\ee
where $\xi < 1$ for $k_T > Q_{s0}$, since $Q_{s0}^2 = A^{1/3}
\Lambda^2 \gg \Lambda^2$. For large transverse momenta in question, $k_T
\gg Q_{s}(y)$, the variable $\xi$ is approaching $1$ from below as
is clear from \eq{xidef}. In the limit $\xi \rightarrow 1$ \eq{rxi}
becomes
\ben
R^{pA} (\xi, y) \bigg|_{\xi \rightarrow 1} \, \approx \, \left( 1 +
(1-\xi ) \frac{\sqrt{y} - \sqrt{Y-y}}{\sqrt{y} + \sqrt{Y-y}} \right) 
\, \exp \left[ - \, 2 \, \sqrt{2 \, \bas \, y \, ( 1-\xi ) \, 
\ln \frac{Q_{s0}}{\Lambda} } \ \right]
\een
\be\label{rxi1}
< \, (2-\xi) \, \exp \left[ - \, 2 \, \sqrt{2 \, \bas \, y \, ( 1-\xi ) \, 
\ln \frac{Q_{s0}}{\Lambda} } \ \right] \, \approx \, 
\exp \left[ - \, 2 \, \sqrt{2 \, \bas \, y \, ( 1-\xi ) \, 
\ln \frac{Q_{s0}}{\Lambda} } \ \right] \, < \, 1, 
\hspace*{.3cm} k_T \gg Q_{s} (y).
\ee
We neglected $2-\xi = 1 + (1-\xi)$ in front of the exponent in
\eq{rxi1} since the $(1-\xi)$ correction to $1$ in it is not enhanced 
by any parametrically large variables, such as $y$ and $\ln Q_{s0} /
\Lambda$ in the exponent. If $R^{pA} (\xi, y)$ from \eq{rxi1} is expanded 
in powers of $(1-\xi)$ this prefactor term would give subleading
logarithmic corrections to the expansion of the exponent, which are
negligible in the DLA limit considered here.

We conclude that $R^{pA}$ from \eq{dlar} is smaller than one.  Since
$R^{pA} (\xi, y)$ in \eq{rxi1} is an increasing function of $\xi$ and
$\xi$ is an increasing function of $k_T$, we observe that $R^{pA} (k_T,
y)$ in \eq{dlar} is an increasing function of $k_T$ approaching $1$
from below. This suppression is mainly due to the difference of the
cutoffs in the logarithms of transverse momentum in the exponent of
\eq{dlar}. The cutoff for the nucleus case is given by nuclear
saturation scale, which is different from the appropriate scale in a
single proton. The high momentum regions, where linear evolution
equations work, are cut off from below by saturation scales, which are
different for different nuclei and for the proton. In this way, as we
can see from Eqs. (\ref{dlar}) and (\ref{rxi1}), saturation influences
the physics at high $k_T$ as long as corresponding $x_{Bj}$ is
small. The effect of saturation is to introduce high-$k_T$
suppression.

The suppression of \eq{rxi1} is a leading twist effect and is due to
quantum evolution. In this sense it is similar to the suppression
suggested in \cite{KLM}. However, the suppression of \cite{KLM}
corresponds to a region of lower $k_T$, where the double logarithmic
approximation of \eq{dlar} is not valid any more. There the
suppression happens due to the change in anomalous dimension of the
gluon distribution function, as we are going to discuss below.


\subsubsection{Onset of Anomalous Dimension: More High-$p_T$ Suppression}

For the values of $k_T$ lower than considered above (but still much
larger than $Q_{s0}$) the saddle point of $\lambda$-integration in
\eq{lt} shifts to a smaller value than given by \eq{spd1}. While in 
determining the saddle point of \eq{spd1} we had to expand $\chi
(\lambda)$ around $\lambda = 2$, now we have to expand it around
$\lambda = 1$. There one writes
\be
\chi (\lambda) \, \approx \, 2 \ln 2 + \frac{7}{4} \, \zeta (3) \, 
(\lambda -1)^2
\ee
obtaining the value of the saddle point
\be\label{sp}
\lambda^*_{sp} \, = \, 1 + \frac{\ln \frac{k_T}{Q_{s0}}}{7 \, \zeta (3) \,
\bas \, y}.
\ee
As was suggested in \cite{IIM1}, the transition of the saddle point
from the value given in \eq{spd1} to the one given in \eq{sp}
happens around 
\be\label{geom}
k_{\mathrm{geom}} \, \approx \, Q_s(y) \, \frac{Q_s (y)}{Q_{s0}} 
\ee
indicating the onset of geometric scaling regime \cite{geom}. Here in
the double logarithmic approximation the saturation scale depends on
energy as \cite{LT,IIM1,Bartels:1992ix}
\be\label{saty}
Q_{s}(y) \, \approx \, Q_{s0} \, e^{2 \, \bas \, y}. 
\ee
The precise value of the scale $k_{geom}$ in \eq{geom} depends on the
definition of the saturation scale and on the way one defines the
transition between the double logarithmic and geometric scaling
regions. For instance, if we define the transition by equating the
saddle points of Eqs. (\ref{spd1}) and (\ref{sp})
\be\label{esp}
\lambda_{sp} = \lambda_{sp}^*, 
\ee
we get at the point of closest approach (the two saddle point values
are never equal to each other)
\be\label{g1}
k_{\mathrm{geom}}\,=\,Q_{s0}\,e^{\bas \, y \ 7^{2/3}\zeta(3)^{2/3}2^{-1/3}}\,
\approx \, Q_{s0} \, e^{3.28 \, \bas \, y}. 
\ee
When combined with the saturation scale from \eq{saty} this gives
\be\label{g2}
k_{\mathrm{geom}} \, \approx \, Q_{s}(y) \, \left( \frac{Q_{s}
(y)}{Q_{s0}} \right)^{0.64},
\ee
which is slightly different from \eq{geom}. At the same time, using
the energy dependence of the saturation scale found in \cite{MT} in
the fixed coupling case
\be\label{mtsat}
Q_{s} (y) \, \approx \, Q_{s0} \, e^{2.44 \, \bas \, y}
\ee
in \eq{g1} gives
\be
k_{\rm geom} \, \approx \, Q_{s} (y) \, \left( \frac{Q_{s}
(y)}{Q_{s0}} \right)^{0.34},
\ee
which is even lower than \eq{g2}. A definition of the transition point
different from \eq{esp} would give slightly different estimates for
$k_{\rm geom}$.

Nevertheless, the ambiguities in the scale $k_{\rm geom}$ notwithstanding,
one can argue, as was done in \cite{IIM1}, that there exists a large
momentum scale $k_{\rm geom}$, which is parametrically larger than the
saturation scale
\be
k_{\rm geom} \, \gg \, Q_{s} (y).
\ee
For $k_T \gsim k_{\rm geom}$ there is no geometric scaling and the gluon
production is well described by double logarithmic approximation
described above resulting in $R^{pA}$ from \eq{dlar}. $k_T \lsim
k_{\rm geom}$ is the region of geometric scaling \cite{IIM1}. When $k_T
\lsim Q_{s} (y)$ (saturation region) multiple pomeron exchanges 
become important leading to the saturation of structure functions
\cite{glr}. For $Q_{s} (y) \lsim k_T \lsim k_{\rm geom}$ (extended
geometric scaling region) multiple pomeron exchanges are not important
yet and the gluon production cross section is described by the leading
twist expression in \eq{lt} with the $\lambda$-integral evaluated near
the saddle point of \eq{sp} \cite{KLM}.

Performing the $\lambda$ and $\lambda'$ integrals in \eq{lt} in the
saddle point approximation around the saddle points of Eqs. (\ref{sp})
and (\ref{spd2}) correspondingly yields
\ben
\frac{d \sigma^{pA \, (1)}}{d^2 k \ dy} \bigg|_{LLA} \, \approx \, 
\frac{C_F \, S_p \, S_A}{\as \, (2 \pi)^4} \
\frac{Q_{s0} \, \Lambda^2}{{\un k}^3} \, \frac{C^A_1}{\sqrt{7 \zeta (3)}}
\, \frac{\ln^{1/4} \frac{k_T}{\Lambda}}{\bas \, (Y-y)^{3/4} \, 
(2 \, \bas )^{1/4}} \, 
\een
\be\label{lla}
\times \, \exp \left[{(\alpha_P - 1) \, y + 2 \, 
\sqrt{2 \, \bas \, (Y-y) \, \ln \frac{k_T}{\Lambda}} - \frac{\ln^2 
\frac{k_T}{Q_{s0}}}{14 \, \zeta (3) \, \bas \, y}}\right],
\ee
where
\be
\alpha_P - 1 \, = \, 2 \, \bas \, \ln 2 
\ee
is the BFKL pomeron intercept \cite{BFKL} and $C^A_1$ is
well-approximated by the first term in the series of \eq{ca}
for all physically reasonable values of $A$
\be\label{ca1}
C^A_1 \, \approx \, \frac{1}{4} \, \left( 1 + \ln
\frac{Q_{s0}}{\Lambda} \right) \, = \,  \frac{1}{4} \, \left( 1 + 
\frac{1}{6} \, \ln A \right).
\ee
The superscript $^{(1)}$ in \eq{lla} denotes the leading twist
contribution. We assume that in the transverse momentum region where
\eq{lla} is valid, $Q_s (y) \lsim k_T \lsim k_{\rm geom}$, the gluon
production cross section in pp collisions is still given by
\eq{dlapp}. This is a good approximation since if $k_T \gsim Q_s (y)
\gg \Lambda$ the double logarithmic approximation of \eq{dlapp} should
work.  Using Eqs. (\ref{lla}) and (\ref{dlapp}) in \eq{rpa} we obtain
\ben
R^{pA} (k_T, y) \bigg|_{Q_s (y) \lsim k_T \lsim k_{\rm geom}} \, = \,
\frac{k_T}{Q_{s0}} \, \frac{2 \, C^A_1}{\sqrt{7
\, \zeta (3)}} \, \frac{\ln^{1/4} \frac{k_T}{\Lambda}}{(2 \, \bas)^{1/4} } 
\, \frac{y^{1/4}}{\sqrt{y} + \sqrt{Y-y}} \, 
\een
\be\label{llar}
\times \, \exp \left[{(\alpha_P - 1) \, y - 2 \, 
\sqrt{2 \, \bas \, y \, \ln \frac{k_T}{\Lambda}} - \frac{\ln^2 
\frac{k_T}{Q_{s0}}}{14 \, \zeta (3) \, \bas \, y}}\right].
\ee
To determine whether $R^{pA} (k_T, y)$ in \eq{llar} is greater or less
than $1$ we first drop the slowly varying and constant prefactors in
front of the exponent and write
\be\label{llar2}
R^{pA} (k_T, y) \bigg|_{Q_s (y) \lsim k_T \lsim k_{\rm geom}} \, \sim \,
\frac{k_T}{Q_{s0}} \, \exp \left[{(\alpha_P - 1) \, y - 2 \, 
\sqrt{2 \, \bas \, y \, \ln \frac{k_T}{\Lambda}} - \frac{\ln^2 
\frac{k_T}{Q_{s0}}}{14 \, \zeta (3) \, \bas \, y}}\right]
\ee
keeping only parametrically important factors. To estimate the value
of $R^{pA}$ in \eq{llar2} in the extended geometric scaling region
$Q_s (y) \lsim k_T \lsim k_{\rm geom}$ we substitute $k_T = k_{\rm
geom}$ into \eq{llar2} with $k_{\rm geom}$ from \eq{geom}. The result
yields an asymptotically small value
\be\label{rpae}
R^{pA} (k_T, y) \bigg|_{Q_s (y) \lsim k_T \lsim k_{\rm geom}} \, \sim \,
e^{- 1.65 \, \bas \, y} \, \ll \, 1,
\ee
where we used $A=197$ for gold nucleus. For other values of $A$ and
for other values of $k_T$ in the region $Q_s (y) \lsim k_T \lsim
k_{\rm geom}$ one still gets exponential suppression for $R^{pA} (k_T, y)$.
Therefore we conclude that $R^{pA} (k_T, y) < 1$ in the extended
geometric scaling region $Q_s (y) \lsim k_T \lsim k_{\rm geom}$.

As can be checked explicitly, for sufficiently large nucleus (large
$A$), $R^{pA} (k_T, y)$ of \eq{llar} is an increasing function of $k$
for $Q_s(y) \lsim k_T \lsim k_{\rm geom}$. As $k_T$ increases it should
smoothly map onto $R^{pA} (k_T, y)$ of \eq{dlar}, which would approach
$1$ from below for asymptotically high $k_T$.

At very high energy the geometric scaling regions for the nucleus
and the proton will overlap. Namely, the geometric scale for the
proton $k_{\rm geom}^p = k_{\rm geom}/A^{1/6}$ will become larger than
the saturation scale for the nucleus $Q_s (y)$ allowing for a region
of $k_T$ where anomalous dimension sets in for gluon production both
in $pA$ and $pp$. \footnote{The onset of anomalous dimension does not
imply saturation and is still a leading twist effect. Therefore
\eq{paevc} in which {\it no} saturation in the proton's wave function was 
assumed is still valid in this region.} In this asymptotic region one
has to estimate the $\lambda$ and $\lambda'$ integrals in \eq{lt}
around the saddle point given by \eq{sp} (with $\Lambda$ instead of
$Q_{s0}$ for the $\lambda'$ integral). Replacing \eq{dlapp} by the
appropriate expression where the saddle points of $\lambda$ and
$\lambda'$ integrals were given by \eq{sp} with $\Lambda$ instead of
$Q_{s0}$ we obtain the following asymptotic expression at mid-rapidity
($y=Y/2$)
\be\label{anas}
R^{pA} (k_T, y) \bigg|_{Q_s (y) \lsim k_T \lsim k_{\rm geom}^p} \, \sim
\, A^{-1/6} \, \exp \left[ \frac{\ln^2 \frac{k_T}{\Lambda} - \ln^2
\frac{k_T}{Q_{s0}}}{14 \, \zeta (3) \, \bas \, y} \right].
\ee
From \eq{anas} we conclude that in the extended geometric scaling
region at asymptotically high energies, $R^{pA}$ saturates to a
parametrically small lower bound, $R^{pA} \sim A^{-1/6}$, which is
independent of energy and is a decreasing function of $A$, or,
equivalently, centrality.

To conclude our discussion of high-$k_T$ suppression at the leading
twist level we note that, as was recently argued in \cite{Mueller3},
the running coupling effects in the BFKL evolution may modify the
$A$-dependence of the saturation scale given by Eqs. (\ref{saty}) and
(\ref{mtsat}), making $Q_s (y)$ almost independent of $A$ at very high
energy corresponding to large rapidity $y$. This would result in
high-$k_T$ suppression which would not disappear at any $k_T$. That is
$R^{pA} (k_T, y)$ would not approach $1$ anymore at high
$k_T$. Instead one would have $R^{pA} (k_T, y) \sim
A^{-1/3}$. \footnote{The argument presented in this paragraph is due
to Larry McLerran.}


\subsection{Next-To-Leading Twist}

Above we have shown that the effect of quantum evolution (\ref{eqN})
on the leading twist gluon production cross section in pA with $k_T >
Q_s (y)$ is to introduce strong suppression of $R^{pA}$. Here we would
like to study the effect of evolution on the gluon production at the
next-to-leading twist level. Below we are going to show that if one
includes the evolution of \eq{eqN} into the next-to-leading twist
correction to \eq{lt} it would start contributing towards enhancement
of $R^{pA}$ at high-$k_T$. This appears to indicate that multiple
rescatterings always tend to enhance gluon production at
high-$k_T$. As we will argue later the effect of quantum evolution is
much stronger. It dominates at high energies leading to overall
suppression of $R^{pA}$. 

A perturbative solution of \eq{eqN} was constructed in \cite{yuri2}
giving the forward amplitude of a $q\bar q$ dipole scattering on the
nucleus as an expansion in powers of $r_T Q_s (y)$
\be\label{exp}
N ({\un r}, {\un b}, y) \, = \, N_1 ({\un r}, {\un b}, y) + 
N_2 ({\un r}, {\un b}, y) + \ldots ,
\ee
where the leading behavior of the $n$th term in the series is $N_n
({\un r}, {\un b}, y) \sim (r_T Q_s (y))^n$. To find the
next-to-leading twist correction to the forward scattering amplitude
of a {\sl gluon} dipole $N_G$ we substitute \eq{exp} into \eq{nga}
obtaining
\be\label{gexp}
N_G ({\un r}, {\un b}, y) \, = \, 2 \, N_1 ({\un r}, {\un b}, y) 
+ 2 \, N_2 ({\un r}, {\un b}, y) - [N_1 ({\un r}, {\un b}, y)]^2
+ \ldots ,
\ee
where the first term on the right is the leading twist contribution
$N_{G \, 1} = 2 \, N_1$ given by \eq{ltNG}, and the next two terms
shown in \eq{gexp} are the next-to-leading twist corrections. Higher
twists are not shown in \eq{gexp}. To calculate the next-to-leading
twist correction to gluon forward amplitude
\be\label{t2}
N_{G \, 2} ({\un r}, {\un b}, y) \, = \, 2 \, N_2 ({\un r}, {\un
b}, y) - [N_1 ({\un r}, {\un b}, y)]^2 \, = \, 2 \, N_2 ({\un r},
{\un b}, y) -
\frac{1}{4} \, [N_{G \, 1} ({\un r}, {\un b}, y)]^2 
\ee
we use $N_{G \, 1}$ from \eq{ltNG} and $N_2$ calculated in
\cite{yuri2}. Employing Eq. (23) from \cite{yuri2} in Eq. (9a) from the 
same reference would give us the first term on the right hand side of
\eq{t2}. In the end we write for a cylindrical nucleus
\ben
N_{G \, 2} ({\un r}, y) \, = \, - \frac{1}{4} \,
\int \frac{d \lambda_1 \, d \lambda_2}{ (2 \pi i)^2} \, 
C_{\lambda_1}^A \, C_{\lambda_2}^A \, (r_T \, Q_{s0})^{\lambda_1 +
\lambda_2} \, e^{2 \, \bas \, y \, [\chi (\lambda_1) + \chi
(\lambda_2)]} \, 
\een
\be\label{ngt2}
\times \left( 2^{\frac{\lambda_1 + \lambda_2}{2}} \, 
\frac{\Gamma\left(\frac{\lambda_1}{2}\right) \, 
\Gamma\left(\frac{\lambda_2}{2}\right) \, \Gamma\left(1 - 
\frac{\lambda_1 + \lambda_2}{2}\right)}{\Gamma\left(1 - 
\frac{\lambda_1}{2}\right) \, \Gamma\left(1 - 
\frac{\lambda_2}{2}\right) \, \Gamma\left( 
\frac{\lambda_1 + \lambda_2}{2}\right)} \, \frac{1}{2 [\chi (\lambda_1) + \chi
(\lambda_2) - \chi (\lambda_1 + \lambda_2)]} + 1 \right).
\ee
The slight difference between the factors in the integrands of
\eq{ngt2} and Eq. (23) of \cite{yuri2} is due to different 
definitions of the coefficients $C^A_\lambda$ (cf. Eq. (15) of
\cite{yuri2} with our \eq{ltNG}).

\begin{figure}
\begin{center}
\epsfxsize=8cm
\leavevmode
\hbox{\epsffile{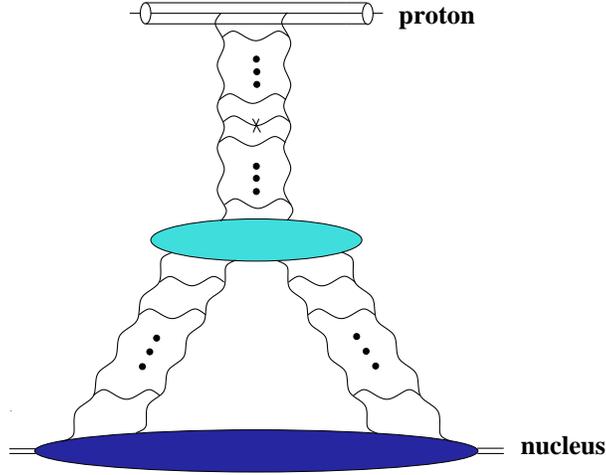}}
\end{center}
\caption{Gluon production in pA collisions at the next-to-leading twist level 
(see text). The blob in the center indicates a triple-pomeron vertex.}
\label{inclht}
\end{figure}

The first term in the parentheses of \eq{ngt2} corresponds to the
first term on the right hand side of \eq{t2}. When we will substitute
$N_{G \, 2}$ from \eq{ngt2} into formula (\ref{paevc}) for the cross
section, this term would give the contribution illustrated in
\fig{inclht}. It corresponds to the case when the gluon is produced 
still by the linear evolution with the triple pomeron vertex inserted
below the emitted gluon. The rapidity of the triple pomeron vertex was
integrated over in arriving at \eq{ngt2}, with only the dominant
contribution corresponding to the vertex being right next to the
emitted gluon left \cite{yuri2}. (As was shown in \cite{KT} the
diagrams where the triple pomeron vertex is inserted above the
produced gluon cancel in the dipole evolution case considered here
\cite{dip,yuri1} in agreement with the expectation of the AGK cutting 
rules \cite{agk}.) The second term in the parenthesis of \eq{ngt2} and
on the right hand side of \eq{t2} corresponds to the case where the
pomeron splitting occurs precisely at the rapidity position of the
gluon production. The emitted gluon is produced by the first step of
the non-linear evolution.  (This term is the main difference between
the results of \cite{KT} and \cite{Braun2}.) As can be seen in the
estimates performed below, this term contributes $50 \div 100 \%$ of
the answer depending on the $k_T$-region in question.

Substituting \eq{ngt2} in \eq{paevc} and integrating over ${\un z}$
yields the following contribution to the gluon production cross
section at the subleading twist level
\ben
\frac{d \sigma^{pA}}{d^2 k \ dy} \bigg|_{SLT} =  - 
\frac{C_F \, S_p \, S_A}{\as \, 2 \, (2 \pi)^3} \,
\int \, \frac{d \lambda_1 \, d \lambda_2 \, d \lambda'}{ (2 \pi i)^3} 
\, C_{\lambda_1}^A \, C_{\lambda_2}^A \, C_{\lambda'}^p \,
\left( \frac{Q_{s0}}{k_T}\right)^{\lambda_1 +
\lambda_2} \, \left(\frac{\Lambda}{k_T}\right)^{\lambda'} 
\een
\ben
\times \,  e^{2 \, \bas \, y \, [\chi (\lambda_1) + \chi (\lambda_2)] +
2 \, \bas \, (Y-y) \, \chi (\lambda')} \
2^{\lambda_1 + \lambda_2 + \lambda' - 3} \ 
\frac{\Gamma\left( \frac{\lambda_1 + \lambda_2 + \lambda'}{2} - 1\right)}{
\Gamma\left( 2 - \frac{\lambda_1 + \lambda_2 + \lambda'}{2} \right)} \,
(\lambda_1 + \lambda_2)^2 \, \lambda'^2
\een
\be\label{slt}
\times \, \left( 2^{\frac{\lambda_1 + \lambda_2}{2}} \, 
\frac{\Gamma\left(\frac{\lambda_1}{2}\right) \, 
\Gamma\left(\frac{\lambda_2}{2}\right) \, \Gamma\left(1 - 
\frac{\lambda_1 + \lambda_2}{2}\right)}{\Gamma\left(1 - 
\frac{\lambda_1}{2}\right) \, \Gamma\left(1 - 
\frac{\lambda_2}{2}\right) \, \Gamma\left( 
\frac{\lambda_1 + \lambda_2}{2}\right)} \, \frac{1}{2 [\chi (\lambda_1) + \chi
(\lambda_2) - \chi (\lambda_1 + \lambda_2)]} + 1 \right).
\ee
To study the onset of higher twist effects, we are interested in the
next-to-leading twist contribution (\ref{slt}) in the region of
transverse momenta $k_T \gsim k_{\rm geom}$. Performing $\lambda_1$
and $\lambda_2$ integrations in \eq{slt} around the saddle point
(\ref{spd1}) and performing the $\lambda'$ integral in
\eq{slt} around the saddle point (\ref{spd2}) yields
\ben
\frac{d \sigma^{pA \, (2)}}{d^2 k \ dy} \bigg|_{DLA} \, \approx \, 
\frac{C_F \, S_p \, S_A \, \sqrt{\pi}}{\as \, (2 \pi)^5} \
\frac{Q_{s0}^4 \, \Lambda^2}{{\un k}^6} \, 
\frac{\ln^{1/4} \frac{k_T}{\Lambda} \, \ln^{1/2} \frac{k_T}{Q_{s0}}}
{( 2\, \bas \, y)^{3/2} \, (2 \, \bas \, (Y-y))^{3/4}} \, 
\left( 2 \, \sqrt{\frac{2 \, \bas \, y}{\ln \frac{k_T}{Q_{s0}}}} 
+ \sqrt{\frac{2 \, \bas \, (Y-y)}{\ln \frac{k_T}{\Lambda}}} \right)
\een
\be\label{dla2}
\times \, \exp \left[4 \, 
\sqrt{2 \, \bas \, y \, \ln \frac{k_T}{Q_{s0}}} + 2 \, 
\sqrt{2 \, \bas \, (Y-y) \, \ln \frac{k_T}{\Lambda}} \right].
\ee
As one can see from \eq{dla2} the next-to-leading twist correction
tends to increase gluon production cross section at high-$k_T$. In the
region of $k_T$ where \eq{dla2} applies, $k_T \gsim k_{\rm geom}$, the
higher twist corrections are parametrically small and can not change
the leading twist suppression of \eq{rxi}. 

To study higher twist effects in the extended geometric scaling
region, $Q_s (y) \lsim k_T \lsim k_{\rm geom}$, we evaluate
$\lambda_1$ and $\lambda_2$ integrals in \eq{slt} around the saddle
point (\ref{sp}) and do the $\lambda'$ integral around the saddle
point (\ref{spd2}) obtaining
\ben
\frac{d \sigma^{pA \, (2)}}{d^2 k \ dy} \bigg|_{LLA} \, \approx \, 
\frac{2 \, C_F \, S_p \, S_A \, \sqrt{\pi}}{\as \, (2 \pi)^5} \
\frac{Q_{s0}^2 \, \Lambda^2}{{\un k}^4} \, 
\frac{(C^A_1)^2}{7 \zeta (3)}
\, \frac{\ln^{1/4} \frac{k_T}{\Lambda}}{\bas \, y \, 
(2 \, \bas \, (Y-y))^{3/4}} \, \left( \sqrt{\frac{2 \, 
\bas \, (Y-y)}{\ln \frac{k_T}{\Lambda}}}
-  \frac{2 \, \ln 
\frac{k_T}{Q_{s0}}}{7 \, \zeta (3) \, \bas \, y} \right)
\een
\be\label{lla2}
\times \, \exp \left[{2 \, (\alpha_P - 1) \, y + 2 \, 
\sqrt{2 \, \bas \, (Y-y) \, \ln \frac{k_T}{\Lambda}} - \frac{2 \, \ln^2 
\frac{k_T}{Q_{s0}}}{14 \, \zeta (3) \, \bas \, y}}\right].
\ee
The sign of \eq{lla2} is determined by the sign of the expression in
the parenthesis. One can see that for very large $k_T$ the expression
in the parenthesis can become negative making the overall contribution
to the cross section negative. However, \eq{lla2} is valid only for
$k_T \lsim k_{\rm geom}$ and, therefore, can not be used at arbitrary
high transverse momenta. At lower $k_T$ the sign changes and the term
in \eq{lla2} begins to contribute toward enhancement of $R^{pA}$. The
value of $k_T$ at which the sign transition takes place depends on the
rapidity in question as well as on the atomic number $A$ of the
nucleus. To estimate the transition value of $k_T$ one has to equate
two terms in the parenthesis of \eq{lla2}. Assuming that $\ln
k_T/Q_{s0} \gg \ln Q_{s0}/\Lambda$ we obtain
\be
k_0 \, \approx \, k_{\rm geom} \, \left( \frac{\Lambda}{Q_{s0}}
\right)^{1/3},
\ee
with $k_{\rm geom}$ given by \eq{g1}. Therefore, the transition from
suppression to enhancement in \eq{lla2} happens at $k_0$ which is
smaller than $k_{\rm geom}$ only by a factor of $A^{-1/18}$, which
indicates that the term in \eq{lla2} is positive inside most of the
extended geometric scaling region contributing to enhancement of gluon
production. Here again one should note that
\eq{lla2} gives us a subleading twist contribution which is
parametrically smaller than the leading twist term from \eq{lla} in
the $k_T$ region at hand ($Q_s (y) \lsim k_T \lsim k_{\rm
geom}$). \eq{lla2} is thus unlikely to affect the suppression of
$R^{pA}$ observed at the leading twist level in Eqs. (\ref{rpae}) and
(\ref{anas}).

We conclude by observing that even after inclusion of quantum
evolution (\ref{eqN}) in the gluon production cross section, multiple
rescatterings (higher twists) still tend to enhance gluon production
at high-$k_T$. In the $k_T$ region considered above, $k_T > Q_s (y)$,
these higher twist effects are still parametrically small. In the next
Subsection we are going to study the region of $k_T$ where all twists
become important, $k_T \sim Q_s (y)$. We will show that the combined
effect of all twists is to introduce suppression of the Cronin
maximum.  


\subsection{Flattening of the Cronin peak}

We have demonstrated that the effect of quantum evolution (\ref{eqN})
is to introduce suppression of $R^{pA} (k_T, y)$ for $k_T
\gsim Q_s (y)$ at the leading twist level. Let us now study 
what happens to $R^{pA} (k_T, y)$ at $k_T \simeq Q_s (y)$ as a result
of evolution in energy. We showed in Sect.~II that in the
quasi-classical approximation the Cronin maximum of the ratio
$R^{pA}(\un k,y)$ occurs at $k_T\simeq Q_{s0}$. In this Subsection we
will follow the value of the ratio $R^{pA}(k_T = Q_s,y)$ to higher
energies when quantum evolution is important. Since the position of
the Cronin maximum is likely to be at $k_T \simeq Q_s (y)$ even when
evolution is included, by studying $R^{pA}(k_T = Q_s,y)$ we are going
to study the dependence of the height of the Cronin maximum on
energy/rapidity.

The fact that the scattering amplitude is a constant at the saturation
scale \cite{Bartels:1992ix,LT,MT} makes our calculation pretty
straightforward. First we assume that Mellin transform of the gluon
dipole amplitude obtained from the \emph{exact} solution to the
evolution equation \eq{eqN} via \eq{nga} can be written as
\be\label{exmel}
N_G(\un z,y)=\int \frac{d\lambda}{2\pi i}\, \left\{ \begin{array}{c}
{\tilde C}_\lambda^A \left[ z_T \, Q_s (y) \right]^\lambda, 
\hspace*{1cm} z_T > \frac{1}{k_{\rm geom}}; \\ \\ 
C_\lambda^A \left( z_T \, Q_{s0}
\right)^\lambda \, e^{2 \, \bas \, \chi (\lambda) \, y}, 
\hspace*{1cm} z_T < \frac{1}{k_{\rm geom}}. \end{array} \right.
\ee
The form of the solution presented in \eq{exmel} assumes geometric
scaling of $N_G$ down to $z_T \simeq 1/k_{\rm geom}$ and a leading twist
expression for smaller $z_T$ in agreement with the analyses of
\cite{IIM1,MT}. Throughout this Subsection we will use the definition 
of the saturation scale $Q_s (y)$ from \eq{saty} and the definition of
$k_{\rm geom}$ from \eq{geom}. Our physical conclusions will be
independent of the choice of definitions for saturation and geometric
scales.

Note that all information about the nonlinear evolution (\ref{eqN}) is
encoded in the function $\tilde C_\lambda^A$ in \eq{exmel}. Using
\eq{exmel} in \eq{paevc} we can calculate the  differential $pA$ 
gluon production cross section at $k_T=Q_s(y)$. Since $k_{\rm geom}
(y) \gg Q_s (y)$ for large $y$ we can set $k_{\rm geom} \rightarrow 
\infty$ neglecting the $z_T < 1/k_{\rm geom}$ part of the integral in
\eq{paevc}. This approximation is justified in the Appendix A. We also assume 
that the dipole amplitude on a proton $n_G$ is still given by the
leading twist expression (\ref{ltnG}) around $k_T \simeq Q_s (y)$,
which is a good approximation for a reasonable size
nucleus. Substituting the first line of \eq{exmel} and \eq{ltnG} into
\eq{paevc} we have
\begin{eqnarray}
\frac{d\sigma^{pA}}{d^2k \, dy}\bigg|_{k_T=Q_s (y)}&=&
 \frac{C_F\, S_p\, S_A \, \Lambda^2}{\as \pi (2\pi)^2}
\int_0^\infty dz_T \,z_T \,J_0(Q_s(y) \, z_T)\int\,\frac{d\lambda}{2\pi i} \, 
\frac{d\lambda'}{2\pi i} \, C_{\lambda'}^p \, \tilde C_\lambda^A \,
\lambda^2 \, \lambda'^2  \nonumber\\ &&\times \, 
(z_T \, \Lambda)^{\lambda'-2}\, [z_T\, Q_s(y)]^{\lambda-2}\,
\, e^{2\,\bas(Y-y)\chi(\lambda')}.
\label{f1}
\end{eqnarray}
Performing the $z_T$-integration in \eq{f1} yields
\ben
\frac{d\sigma^{pA}}{d^2k \, dy}\bigg|_{k_T=Q_s (y)} \, = \,
 \frac{C_F\, S_p\, S_A }{\as \pi (2\pi)^2} \int\,\frac{d\lambda}{2\pi i} \, 
\frac{d\lambda'}{2\pi i} \, C_{\lambda'}^p \, \tilde C_\lambda^A \,
\lambda^2 \, \lambda'^2 \, 2^{\lambda + \lambda' -3} \,  
\frac{\Gamma \left( \frac{\lambda + \lambda'}{2} - 1 \right)}{\Gamma 
\left( 2 - \frac{\lambda + \lambda'}{2} \right)} 
\een
\be\label{fm} 
\times \, \left( \frac{\Lambda}{Q_s (y)} \right)^{\lambda'}
e^{2\,\bas(Y-y)\chi(\lambda')}.  
\ee
It can be seen that all energy/rapidity and almost all atomic number
dependence in \eq{fm} is given by the $\lambda'$-integral. Since $Q_s
(y) \gg \Lambda$, the integral over $\lambda'$ in \eq{fm} can be
evaluated in the double logarithmic approximation around the saddle
point of \eq{spd2} taken at $k_T = Q_s (y)$. After that the integral
over $\lambda$ carries almost no dynamical information.  The result
reads
\be\label{f2}
\frac{d\sigma^{pA}}{d^2k \, dy}\bigg|_{k_T=Q_s (y)} \, = \,
\frac{C_F\, S_p\, S_A }{\as \, (2\pi)^4} \, \sqrt{\pi} \, 
{\cal C}_A \, \frac{\ln^{1/4} \frac{Q_s (y)}{\Lambda}}{[2 \,
\bas \, (Y-y)]^{3/4}} \, \frac{\Lambda^2}{Q_s^2 (y)} \, \exp \left( 2
\sqrt{2 \, \bas
\, (Y-y) \, \ln \frac{Q_s (y)}{\Lambda}} \right),
\ee
where the integration over $\lambda$ gave an unknown function
${\cal C}_A$ defined as
\be\label{clam}
{\cal C}_A \, = \, \int \frac{d\lambda}{2\pi i} \, \lambda^2 \, 
\tilde C_\lambda^A \, 2^{\lambda} \, \frac{\Gamma \left( 
\frac{\lambda}{2} \right)}{\Gamma \left( 1 - \frac{\lambda}{2} \right)}.  
\ee
Here we assume that ${\cal C}_A$ is only weakly (at most
logarithmically) dependent on $A$, as is true for other coefficients
like the one shown in \eq{ca1}. In case of an ``ideal'' geometric
scaling the coefficient $C_\lambda^A$ in \eq{exmel} would be
completely $A$-independent ridding ${\cal C}_A$ of all of its
$A$-dependence as well.

To construct $R^{pA}$ we take the gluon production cross section in
$pp$ from \eq{dlapp} putting $k_T = Q_s (y)$. Substituting
Eqs. (\ref{f2}) and (\ref{dlapp}) into \eq{rpa} we get
\be\label{f3}
R^{pA}(Q_s (y),y) \, = \, \sqrt{\pi} \, {\cal C}_A \, (2 \,
\bas)^{1/4}\, \frac{y^{3/4}}{\sqrt{y}+\sqrt{Y-y}}\,
\ln^{1/4} \left( \frac{Q_s(y)}{\Lambda} \right) \,
\frac{Q_s^2(y)}{Q_{s0}^2}\, \exp \left( - 2 \, 
\sqrt{2 \, \bas \, y \, \ln \frac{Q_s(y)}{\Lambda}} \right).
\ee
The energy and $A$-dependence of \eq{f3} can be found using \eq{saty}
and keeping in mind that $Q_{s0}=A^{1/6} \, \Lambda$. Since the
definition of the saturation scale (\ref{saty}) is valid up to
logarithmic prefactors, we can drop the prefactors in \eq{f3} leaving
only
\be\label{f4}
R^{pA}(Q_s (y),y) \, \propto \, \frac{Q_s^2(y)}{Q_{s0}^2}\,
\exp \left( - 2 \, \sqrt{2 \, \bas \, y \, \ln \frac{Q_s(y)}{\Lambda}} \right).
\ee
(If one defines $Q_s (y)$ by taking $N_G (z_T = 1/Q_s (y), y)$ in the
double logarithmic approximation and requiring that $N_G (z_T = 1/Q_s
(y), y) =$const \cite{LT,Bartels:1992ix,MT} the prefactors in \eq{f3}
would cancel exactly.) Using \eq{saty} in \eq{f4} yields
\be\label{f5}
R^{pA}(Q_s (y),y) \, \propto \, \exp\left\{4\, \bas \,y\,
\left(1-\sqrt{1+\frac{\ln A^{1/6}}{2\,\bas \, y}}\right)\right\} \, < \, 1. 
\ee
We observe from \eq{f5} that in the course of quantum evolution the
Cronin maximum of the ratio $R^{pA}$ decreases with energy until, at
very high energy, it saturates at the lowest value $R^{pA} \sim
A^{-1/6}$, which is much less than $1$. The height of the Cronin peak
is also a decreasing function of collision centrality/atomic number
$A$, as can be seen from \eq{f5} \footnote{We have recently learned
that a similar conclusion regarding centrality dependence of the
Cronin peak has been reached by A. Mueller and collaborators.}.

Applicability of \eq{f5} is restricted by applicability of
Eqs. (\ref{dlapp}) and (\ref{f2}). The latter two equations are valid
only in the region where $Q_s (y)$ is larger than the geometric scale
of the proton: $Q_s (y) > k_{\rm geom}/A^{1/6}$ with $k_{\rm geom}$
taken from \eq{g1}, since this is where the transition between the
saddle points takes place. With the help of \eq{geom} this condition
becomes $Q_s (y) < A^{1/6} Q_{s0}$. In the kinematic region of
extremely high $y$ where this condition is not satisfied anymore, one
has to replace Eqs. (\ref{f2}) and (\ref{dlapp}) by appropriate cross
sections evaluated around the saddle point of \eq{sp}. To generalize
the conclusions presented above to arbitrary high rapidity let us
follow Mueller and Triantafyllopoulos \cite{MT} and define the
saturation scale by requiring that the power of the exponent in the
leading twist expression for $N_G$ given by (cf. \eq{ltNG})
\ben
N_{G \, 1} ({\un z}, y) \, = \, \int \frac{d \lambda}{2 \pi i} \,
C_\lambda^A \, e^{2 \, \bas \, \chi (\lambda) \, y +
\lambda \, \ln (z_T \, Q_{s0})}
\een
is zero and stationary (its derivative with respect to $\lambda$ is
zero) at $z_T = 1/Q_s (y)$. These conditions are satisfied at
$\lambda_0 = 1.255$ \cite{MT} (our definition of $\chi (\lambda)$ is
different from the one used in \cite{MT}). Resulting saturation scale
is given by \eq{mtsat}. Arguing that the $\lambda$-integral in
\eq{ltNG} is dominated by the saddle point in the exponent we conclude 
that at $z_T = 1/Q_s (y)$ the gluon dipole amplitude $N_G (z_T = 1/Q_s
(y), y)$ is approximately constant at large $y$ \cite{MT}
\be\label{const}
 \int \frac{d \lambda}{2 \pi i} \, C_\lambda^A \, \left(
 \frac{Q_{s0}}{Q_s (y)} \right)^\lambda \, e^{2 \, \bas \, \chi
 (\lambda) \, y} \, = \, \int \frac{d \lambda}{2 \pi i} \,
C_\lambda^A \, e^{2 \, \bas \, \chi (\lambda) \, y +
\lambda \, \ln (Q_{s0} / Q_s (y))} \, \simeq \, \mbox{constant} (y, A).
\ee 
Making a similar assumption about the $\lambda'$-integral in \eq{fm}
taken at mid-rapidity ($y = Y/2$) and remembering that $Q_{s0} =
A^{1/6} \, \Lambda$ yields
\be\label{s1}
\frac{d\sigma^{pA}}{d^2k \, dy}\bigg|_{k_T=Q_s (y), \, y=Y/2} \, 
\sim \, S_A \, A^{- \lambda_0 /6} \, \sim \, A^{2/3 
- \lambda_0 /6}.
\ee
Modifying \eq{lt} to give gluon production in $pp$ at $k_T = Q_s (y)$
also taken at mid-rapidity we obtain
\ben
\frac{d \sigma^{pp}}{d^2 k \ dy} \bigg|_{k_T=Q_s (y), \, y=Y/2} \, = \, 
\frac{C_F \, S_p^2}{4 \, \as \, (2 \pi)^3} \, 
\int \frac{d \lambda}{2 \pi i} \, \frac{d \lambda'}{2 \pi i} \,
\lambda^2 \, \lambda'^2 \, C_\lambda^A \, C_{\lambda'}^p \, 
2^{\lambda + \lambda'} \, \frac{\Gamma \left( -1 + \frac{\lambda +
\lambda'}{2} \right)}{\Gamma \left( 2 - \frac{\lambda +
\lambda'}{2} \right)}  
\een
\be\label{ppqs}
\times \, \left( \frac{\Lambda}{Q_s (y)} \right)^{\lambda + \lambda'} \,  
e^{2 \, \bas \, [\chi (\lambda) + \chi (\lambda')] \, y}.
\ee
Again the $\lambda$- and $\lambda'$-integrals in \eq{ppqs} are
dominated by the saddle points at $\lambda_0$ giving an
energy-independent cross section scaling as
\be\label{s2}
\frac{d \sigma^{pp}}{d^2 k \ dy} \bigg|_{k_T=Q_s (y), \, y=Y/2} \, 
\sim \,  A^{- 2 \, \lambda_0 /6}
\ee
with atomic number $A$. Combining Eqs. (\ref{s1}) and (\ref{s2}) with
\eq{rpa} yields
\be\label{s3}
R^{pA} (Q_s (y),y) \, \propto \, A^{-1/3 + \lambda_0 /6} \,
\mbox{constant} (y) \, \sim \, A^{- 0.124},
\ee
for $\lambda_0 = 1.255$. Note that the power of $A$ in \eq{s3} is
pretty close to that following from \eq{f5} and the two powers would
be identical for $\lambda_0 = 1$. Note also that taking the expression
for $R^{pA}$ in the geometric scaling region from \eq{anas} and
extrapolating it down to $k_T = Q_s(y)$ one would obtain a power of
$A$ very close to that in \eq{s3} if one uses $Q_s(y)$ from
\eq{saty}. This conclusion not only verifies the self-consistency 
of our analysis, but also demonstrates that at asymptotic energies the
height of Cronin maximum becomes (parametrically) equal to the height
of the rest of the $R^{pA}$ curve in the extended geometric scaling
region. This is likely to indicate that at these energies the curve
flattens out and the Cronin peak disappears.

With the help of \eq{s3} we conclude that at high rapidities/energies
the Cronin maximum decreases with energy and centrality, with $R^{pA}
(Q_s (y),y)$ becoming less than $1$.  Eventually, at very high energy,
the Cronin peak flattens out and saturates to an energy independent
lower limit given by \eq{s3}, which is parametrically suppressed by
powers of $A$.


\subsection{Suppression Deep Inside Saturation Region}

Above we have shown that non-linear evolution (\ref{eqN}) introduces
suppression of gluon production in $pA$ collisions making $R^{pA} < 1$
for $k_T \gsim Q_s (y)$. In the region of smaller $k_T$,  $k_T \ll Q_s (y)$, 
we observed in Sect. IIIB that in the quasi-classical case of
McLerran-Venugopalan model the ratio $R^{pA} \ll 1$ (see
\eq{rpa2}). When the quantum evolution (\ref{eqN}) is included it makes 
sense to consider the interval of low $k_T$ bounded from below by the
saturation scale of the proton $\Lambda_s (y)$, such that $\Lambda_s
(y) \ll k_T \ll Q_s (y)$. (For $k_T \lsim \Lambda_s (y)$ the proton
wave function also saturates and particle production in both $pp$ and
$pA$ becomes similar to the case of $AA$, which has not been resolved
even at the quasi-classical level \cite{KV,lappi,yuriaa}. Inclusion of
evolution in $AA$ is an even more difficult problem which we are not
going to address here.) If $k_T$ is larger than the geometric scale of
the proton $k_{\rm geom}/A^{1/6}$ (but still much less than $Q_s (y)$)
we can use \eq{dlapp} to describe the gluon production cross section
in $pp$. Deep inside the saturation region in $pA$ the gluon
production has been estimated in
\cite{KT}. Employing Eq. (57) from \cite{KT} together with \eq{dlapp}
we conclude that at mid-rapidity
\be\label{rpalqs}
R^{pA} (k_{\rm geom}^p < k_T \ll Q_s (y),y) \, \sim \,
\frac{k_T^2}{Q_{s0}^2} \, e^{- 2
\sqrt{2 \, \bas \, y \, \ln k_T/\Lambda}}.
\ee
\eq{rpalqs} shows that inclusion of quantum evolution only introduces 
more suppression into $R^{pA}$ at $k_T \ll Q_s (y)$, making it a
decreasing function of both the atomic number and energy. At very high
energy $k_T$ may become smaller than the geometric scale for the
proton $k_{\rm geom}/A^{1/6}$ and the gluon production in $pp$ would
be driven by the saddle point (\ref{sp}) with $\Lambda$ instead of
$Q_{s0}$. Similarly to how it was done in \cite{KT} for DLA, one can
estimate the gluon production cross section (\ref{paevc}) deep inside
the saturation region with the dipole amplitude on the proton
evaluated around the LLA saddle point $\lambda' \approx 1$. The result
at mid-rapidity yields
\be\label{s6}
R^{pA} (\Lambda_s (y) \ll k_T \le k_{\rm geom}^p) \, \sim \, A^{-1/3}
\, \frac{Q_s (y)}{\Lambda} \, \exp \left[ - (\alpha_P - 1) y + \frac{2
\, \ln^2 (k_T/\Lambda) - \ln^2 (Q_s(y)/\Lambda)}{14 \, \zeta (3) \, \bas \, y} 
\right].
\ee
Therefore, at very high energies the ratio $R^{pA}$ becomes almost
independent of $k_T$ even at very low $k_T$. Using the saturation
scale from \eq{saty} in \eq{s6} at asymptotic energies gives
\be\label{s7}
R^{pA} (\Lambda_s (y) \ll k_T \le k_{\rm geom}^p) \, \sim \,
A^{-0.2} \, e^{- 1.0 \, \bas \, y}.
\ee
We observe again that nonlinear evolution leaves $R^{pA}$ very small
at $k_T \ll Q_s(y)$. $R^{pA}$ given by \eq{s7} is a decreasing
function of both rapidity/energy and centrality. This conclusion seems
natural, since the saturation effects are known to soften the
low-$k_T$ gluon spectra in pA compared to pp.


\subsection{Toy Model}

To illustrate the conclusions reached above let us construct a simple
toy model exhibiting suppression of $R^{pA}$ at all $k_T$. We start
with the quasi-classical formula for gluon production in $pA$ in the
following form which could be obtained from \eq{pamv3} for a
cylindrical nucleus and for azimuthally symmetric $N_G$
\be\label{toyx}
\frac{d \sigma^{pA}}{d^2 k \ dy} \, = \, 
\frac{\as C_F}{\pi^2} \, \frac{S_A}{k_T^2} \, \int_0^\infty \,  d z_T \, 
J_0 (k_T \, z_T) \, \ln \frac{1}{z_T \Lambda} \
\partial_{z_T} \, [z_T \ \partial_{z_T} N_G (z_T, y=0)].
\ee
With the increase of energy the gluon dipole amplitude on the nucleus
will reach saturation. Therefore, its $z_T$-dependence will change
more significantly than for the corresponding amplitude on the proton,
which will stay unsaturated. (Of course at very high energy the dipole
amplitude on the proton will also reach saturation, but we are not
going to consider that energy range here.) Therefore, in our toy model
we will assume for simplicity that the gluon dipole amplitude on the
proton remains unchanged with increasing energy, giving $\ln 1/ (z_T
\Lambda)$ in \eq{toyx}. We will model the gluon dipole amplitude at 
high energy by a Glauber-like unitary expression
\be\label{toyN}
N_G^{toy} (z_T, y) \, = \, 1 - e^{- z_T \, Q_s(y)},
\ee
which mimics the onset of anomalous dimension $\lambda =1$ by the
linear term in the exponent. The saturation scale $Q_s(y)$ in
\eq{toyN} is some increasing function of $y$ which can be taken from 
\eq{saty} or from \eq{mtsat}. Indeed the amplitude in \eq{toyN} has 
an incorrect small-$z_T$ behavior, scaling proportionally to $z_T$
instead of $z_T^2$ as shown in \eq{Ndla}. If \eq{toyN} is used in
\eq{toyx} it would lead to an incorrect high-$k_T$ behavior of the
resulting cross section. We therefore argue that \eq{toyN} is,
probably, a reasonable model for $N_G$ inside the saturation and
extended geometric scaling regions ($1/z_T \sim k_T < k_{\rm geom}$),
but should not be used for very small $z_T$ / high $k_T$ ($1/z_T \sim
k_T > k_{\rm geom}$).

Substituting \eq{toyN} into \eq{toyx} and integrating over $z_T$
yields
\ben
\frac{d \sigma^{pA}_{toy}}{d^2 k \ dy} \, = \, 
\frac{\as C_F}{\pi^2} \, \frac{S_A}{k_T^2} \, \frac{Q_s}{k_T^2 + Q_s^2} 
\, \left[ - \, Q_s (k_T^2 + Q_s^2) + \sqrt{k_T^2 + Q_s^2} \left( 2 Q_s^2 + 
\gamma \, k_T^2 + k_T^2 \, \ln \frac{2 (k_T^2 + Q_s^2)}{k_T \, \Lambda} 
\right. \right.
\een
\be\label{toypA}
\left. \left. + \frac{k_T^2}{2} \, 
\ln \frac{\sqrt{k_T^2 + Q_s^2} - Q_s}{\sqrt{k_T^2 + Q_s^2} 
+ Q_s} \right) \right],
\ee
where $\gamma$ is the Euler's constant and $Q_s = Q_s
(y)$. Corresponding gluon production cross section for $pp$ is
obtained by expanding \eq{toypA} to the lowest order at high $k_T$ and
substituting $\Lambda$ instead of $Q_s$ and $S_p$ instead of $S_A$:
\be\label{toypp}
\frac{d \sigma^{pp}_{toy}}{d^2 k \ dy} \, = \, \frac{\as C_F}{\pi^2} \, 
\frac{S_p \, \Lambda}{k_T^3} \, \left( \ln \frac{2 \, k_T}{\Lambda} + 
\gamma \right).
\ee
Of course in \eq{toypp} one implicitly assumes that anomalous
dimension has set in for only one of the protons in $pp$. This
assumption is not valid at mid-rapidity, but may be used to study
particle production at rapidities near fragmentation region of one of
the protons.

Substituting Eqs. (\ref{toypA}) and (\ref{toypp}) in \eq{rpa} yields
\ben
R^{pA}_{toy} (k_T, y) \, = \, \frac{k_T \, \Lambda}{Q_s \, (k_T^2 +
Q_s^2) \, (\ln (2 \, k_T/\Lambda) + \gamma)}
\, \left[ - \, Q_s (k_T^2 + Q_s^2) + \sqrt{k_T^2 + Q_s^2} \left( 2 Q_s^2 + 
\gamma \, k_T^2 \right. \right.
\een
\be\label{toyrpa}
\left. \left. + k_T^2 \, \ln \frac{2 (k_T^2 + Q_s^2)}{k_T \, \Lambda} 
 + \frac{k_T^2}{2} \, 
\ln \frac{\sqrt{k_T^2 + Q_s^2} - Q_s}{\sqrt{k_T^2 + Q_s^2} 
+ Q_s} \right) \right],
\ee
in which we assumed that $\Lambda$ is the saturation scale of the
proton such that $Q_s^2 = A^{1/3} \, \Lambda^2$ even at high energy.

\begin{figure}
\begin{center}
\epsfxsize=10cm
\leavevmode
\hbox{\epsffile{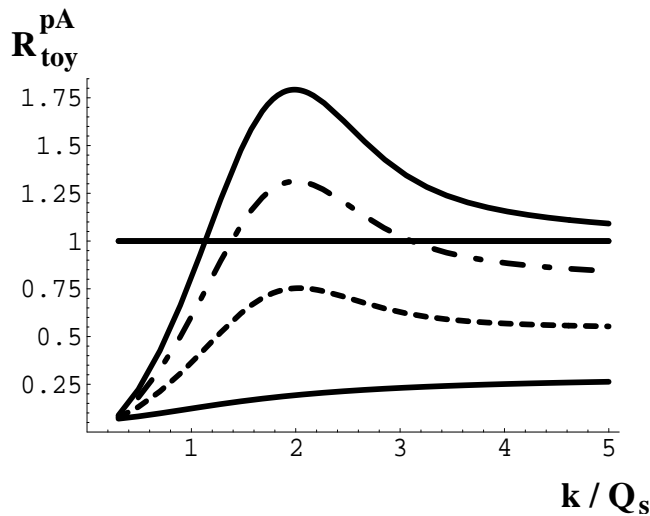}}
\end{center}
\caption{The ratio $R^{pA}$ plotted as a function of $k_T/Q_s$ for 
(i) McLerran-Venugopalan model, which is valid for moderate energies
(upper solid line); (ii) our toy model for very high
energies/rapidities from \eq{toyrpa} (lower solid line); (iii) an
interpolation to intermediate energies (dash-dotted and dashed
lines). The cutoff is $\Lambda = 0.3 \, Q_s$. }
\label{toy}
\end{figure}

The toy model $R^{pA}_{toy} (k_T, y)$ from \eq{toyrpa} is plotted as a
function of $k_T/Q_s$ in \fig{toy} for $\Lambda = 0.3 \, Q_s$ (lower
solid curve). It exhibits suppression of gluon production in $pA$ at
all values of $k_T$ leveling off at $R^{pA}_{toy} \sim \Lambda/Q_s
\sim A^{-1/6}$ for $k_T \gsim Q_s$ at high energy, in agreement with our 
conclusions of Sections IIIB and IIID.

Our toy model (\ref{toyrpa}) represents the high energy asymptotics of
$R^{pA}$.  To compare it to lower energies, we also plot $R^{pA}$ for
the quasi-classical McLerran-Venugopalan model given by
\eq{rpa1} (upper solid curve in \fig{toy}). As the energy increases the 
upper solid line in \fig{toy} would decrease eventually turning into
the lower solid line. The corresponding intermediate energy stages are
shown by the dash-dotted and dashed lines in \fig{toy}. These lines
are for illustrative purposes only and do not correspond to any toy
model. They demonstrate how the Cronin peak gradually disappears as
energy or rapidity increase.


\section{Conclusions}

In this paper we have demonstrated that saturation effects in the
gluon production in pA at moderate energy can be taken into account in
the quasi-classical framework of McLerran-Venugopalan model, which
includes Glauber-Mueller multiple rescatterings, resulting only in
Cronin enhancement of produced gluons at $k_T = (1 \div 2) \, Q_{s0}$,
as was shown in \fig{cron} and in \eq{qclt}.  Similar conclusions have
been reached in \cite{KNST}. In this quasi-classical
approximation the height and position of the Cronin peak are
increasing functions of centrality as indicated by \eq{max}.

We have also shown that at higher energies/rapidities, when quantum
evolution becomes important, it introduces suppression of gluons
produced in pA collisions at all values of $k_T$, as compared to the
number of gluons produced in pp collisions scaled up by the number of
collisions $N_{coll}$, as suggested previously \cite{KLM}. The
resulting $R^{pA}$ at high energy/rapidity is a decreasing function of
centrality. We have considered three different complimentary regions
of $k_T$, which cover together all of $k_T$-range:

\begin{enumerate}

\item[i.] $k_T >  Q_s (y)$ region. Gluon production cross section in $pA$
is dominated by the leading twist effects in this region of $k_T$.  We
have shown how the leading twist suppression arises in the double
logarithmic approximation for $k_T > k_{\rm geom}
\gg Q_s (y)$ with the corresponding $R^{pA} (k_T,y)$ given by
\eq{dlar}, which approaches $1$ as $k_T \rightarrow \infty$. At 
$Q_s (y) < k_T \lsim k_{\rm geom}$ the leading twist suppression is
mainly due to the change in anomalous dimension $\lambda$ from its
double logarithmic value (\ref{spd1}) to the leading logarithmic value
(\ref{sp}). $R^{pA} (k_T,y)$ for this $k_T$-window is given by
\eq{llar} leading to suppression described by 
\eq{rpae}. At very high energies, when the extended geometric scaling
regions of the proton and the nucleus overlap (for $Q_s (y) < k_T
\lsim k_{\rm geom}^p$) the decrease of $R^{pA}$ with energy stops at
roughly $R^{pA} \sim A^{-1/6}$ as follows from \eq{anas}. This leading
twist effect has been originally pointed out in \cite{KLM}. We have
not considered suppression mechanisms that may stem from running of
the coupling constant, which would modify the $A$-dependence of the
saturation scale \cite{Mueller3}.

\item[ii.] $k_T \sim Q_s (y)$ is the position of the Cronin maximum in 
the quasi-classical approximation. We began the analysis of this
$k_T$-region by studying higher twists in the adjacent region of $k_T
> Q_s (y)$. Next-to-leading twist term was shown to contribute towards
enhancement of $R^{pA}$ at high-$k_T$ even when evolution is
included. However, higher twist effects are parametrically small at
$k_T > Q_s (y)$ and can not change our leading twist conclusions about
suppression. To assess the contribution of all twists we studied the
behavior of the Cronin maximum ($k_T \sim Q_s (y)$) with increasing
energy. We showed that $R^{pA}$ at $k_T = Q_s (y)$ is a decreasing
function of energy/rapidity and centrality saturating at the
energy-independent lower bound given by \eq{s3}. Since the height of
the Cronin maximum becomes parametrically of the same order as the
rest of $R^{pA}$ at higher $k_T$ given by \eq{anas}, we conclude that
Cronin peak disappears at asymptotically high energies/rapidities.

\item[iii.] $k_T \ll Q_s (y)$ region. The suppression of $R^{pA}$ deep 
inside the saturation region, $k_T \ll Q_s (y)$, only gets stronger as
the evolution (\ref{eqN}) is included (see \eq{s7}).

\end{enumerate}

Our results are summarized in \fig{toy}.

It is interesting to observe that the behavior of $R^{pA}$ at high
energies is qualitatively different from what one would expect by
taking the quasi-classical expression (\ref{rpa1}) and letting $Q_s$
in it increase with energy. In case of DIS a similar trick where one
replaces $Q_{s0}$ in the Glauber-Mueller expression for the dipole
cross section (\ref{GM}) by the energy dependent $Q_s$ from, for
instance, \eq{saty} leads to correct qualitative behavior of resulting
$F_2$ structure function and even generates some successful
phenomenology \cite{GBW}. However, as we showed above, a naive
generalization of McLerran-Venugopalan model by increasing $Q_s$ with
energy does not work for $R^{pA}$ even at the qualitative level. 

The analysis in the paper was, of course, done for sufficiently high
energy and/or rapidity, such that the saturation approach was assumed
to be still valid for the highest $k_T$ involved. This implies that
the effective Bjorken $x$ is still sufficiently small for all $k_T$ we
consider. The extent to which this treatment applies at high $k_T$
hadron production at RHIC is difficult to assess theoretically. We
thus eagerly await the results of the experimental analyses of
centrality dependence of hadron production above the Cronin region
($k_T \geq 6$ GeV). It is also very important to extend the present
measurements away from the central rapidity region to separate initial
state effects from possible energy loss in cold nuclear
matter. Indeed, in the deuteron fragmentation region, the effects of
saturation in the $Au$ wave function will be enhanced, while the
density of the produced particles (see, e.g., the predictions in
\cite{KLNd}) and thus the associated energy loss will be minimal. In
the $Au$ fragmentation region the opposite will be true.

We, therefore, conclude that if the effects of quantum evolution and
anomalous dimension are observed in the forward rapidity region of
$dAu$ collisions at RHIC, they would manifest themselves by reducing
$R^{dA}$ at all $k_T$ as shown in \fig{toy}, eliminating the Cronin
enhancement.  $R^{dA}$ will become a {\it decreasing} function of
centrality. The $pA$ program at LHC would observe an even stronger
suppression of $R^{pA}$. However, it might be that the quantum
evolution effects are still not important even in the forward region
of $dAu$ collisions at RHIC.  Then reduction of $R^{dA}$ going from
mid-rapidity to deuteron fragmentation region should be rather mild
and the Cronin peak would not disappear in the forward region. The
relevant particle production physics would be described by
McLerran-Venugopalan model. The height of the Cronin peak would then
be an {\it increasing} function of centrality.

If the forthcoming data on $R^{dA}$ in the forward rapidity region of
$dAu$ collisions would have no high-$p_T$ suppression and would
exhibit only a strong Cronin maximum which is an increasing function
of centrality in agreement with predictions of multiple rescattering
models described in Sect. III
\cite{KNST,AG,ktbroadening1,ktbroadening2,Vitev03,ktbroadening3}, 
then all of the observed high-$p_T$ suppression in $Au - Au$
collisions would have to be attributed to the final state
effects. However, if the future $R^{dA}$ data in the forward rapidity
region exhibits suppression either for all $p_T$ or at high $p_T$ with
$R^{dA}$ being a decreasing function of centrality as described in
this paper (see also \cite{KLM}), then a fraction of $R^{AA}$
suppression in the forward rapidity region of $Au - Au$ collisions
should be attributed to initial state quantum evolution
effects. Indeed, there is some evidence \cite{Arsene:2003yk} that the
high $k_T$ suppression in $Au-Au$ collisions increases between the
pseudo--rapidities $\eta =0$ and $\eta = 2.2$.

The $dAu$ data at $y \simeq 0$
\cite{dAtaphen,dAtaphob,dAtastar,Arsene:2003yk} also suggest suppression 
of the yields of charged hadrons \cite{dAtastar} and neutral pions
\cite{dAtaphen} at $k_T \geq 6$ GeV,  though the suppression is 
not significant statistically. If this initial--state effect is
confirmed, it should also be taken into account in the interpretation
of $Au-Au$ results at $y \simeq 0$.

The $dAu$ results will thus allow to clarify the relative importance
of initial and final state interactions at different transverse
momenta and rapidities of the produced particles. They will be
indispensable for establishing a complete physical picture of heavy
ion collisions at RHIC energies.

{\it Note added:} After the first version of this paper appeared, a
similar analysis has been done in \cite{BKW,JNV,AAKSW}. The analyses
of \cite{BKW,JNV,AAKSW} agree with our conclusions on the presence of
Cronin effect in the quasi-classical approximation. The results of
\cite{BKW,AAKSW} are also in agreement with our conclusion about high-$p_T$ 
suppression of {\it gluon} production.

\section*{Acknowledgments} 

We are grateful to Al Mueller for illuminating discussions of the
earlier version of the paper and to him, Eugene Levin and Larry
McLerran for continuing enjoyable collaborations on the subject.  The
authors would like to thank Alberto Accardi, Rolf Baier, Miklos
Gyulassy, Jamal Jalilian-Marian, Alex Kovner, Xin-Nian Wang, Heribert
Weigert and Urs Wiedemann for stimulating and informative discussions.

The research of D. K. was supported by the U.S. Department of Energy
under Contract No. DE-AC02-98CH10886.  The work of Yu. K. was
supported in part by the U.S. Department of Energy under Grant
No. DE-FG03-97ER41014. The work of K. T. was sponsored in part by the
U.S. Department of Energy under Grant No. DE-FG03-00ER41132.


\appendix\section{}

Here we are going to derive \eq{f1}. In writing down \eq{f1} we
assumed that the integration region $z_T<1/k_{\rm geom}$ is
negligible. To justify this approximation let us start by substituting
\eq{exmel} into \eq{paevc}. We find
\begin{eqnarray}
\frac{d\sigma^{pA}}{d^2k \, dy}\bigg|_{k_T=Q_s (y)}&=& \frac{C_F\, S_p\,
  S_A}{\as \pi (2\pi)^2\, Q_s^2 (y)}\,\bigg\{
\int_0^{1/k_{\rm geom}}dz_T \, z_T \,J_0(Q_s(y) z_T )\int\,\frac{d\lambda}{2\pi
  i} \int\,\frac{d\lambda'}{2\pi i}\, C_{\lambda'}^p\, C_\lambda^A\,
\lambda^2\,\lambda'^2\,\Lambda^2 \, Q_{s0}^2 \nonumber\\
&&\times\, (z_T \, \Lambda)^{\lambda'-2}\, (z_T \, Q_{s0})^{\lambda-2}\,
e^{2 \, \bas \, y \, \chi(\lambda)+2 \, \bas \, (Y-y) \,\chi(\lambda')}
\nonumber\\
&&+\int_{1/k_{\rm geom}}^\infty dz_T \, z_T \, J_0(Q_s (y) z_T)
\int\,\frac{d\lambda}{2\pi i} \int\,\frac{d\lambda'}{2\pi i}\, 
C_{\lambda'}^p\, \tilde C_\lambda^A\,
\lambda^2\,\lambda'^2\,\Lambda^2 \, Q_{s}^2 (y) \nonumber\\ 
&&\times\, (z_T \, \Lambda)^{\lambda'-2}\, [z_T \, Q_{s}(y)]^{\lambda-2}\,
e^{2 \, \bas \, (Y-y) \, \chi(\lambda')}\bigg\}. \label{uwa}
\end{eqnarray}
The difference between \eq{uwa} and the target \eq{f1} is
\ben
 \frac{C_F\, S_p\,
  S_A}{\as \pi (2\pi)^2\, Q_s^2 (y)} \,
\int_0^{1/k_{\rm geom}}dz_T \, z_T^{-3} \, J_0(Q_s(y) z_T ) \, \int\,
\frac{d\lambda}{2\pi i} \int\,\frac{d\lambda'}{2\pi i}\, 
C_{\lambda'}^p \, \lambda^2\,\lambda'^2
\een
\be
\times \, (z_T \, \Lambda)^{\lambda'}\,
e^{2 \, \bas \, (Y-y) \, \chi(\lambda')} \, \bigg[ 
C_\lambda^A \, 
 (z_T \, Q_{s0})^{\lambda}\,
e^{2 \, \bas \, y \, \chi(\lambda)}
- \tilde C_\lambda^A \, 
[z_T \, Q_{s}(y)]^{\lambda}\bigg]. \label{diff}
\ee
Since $k_{\rm geom} \gg Q_s(y)$ we can neglect the argument of the
Bessel function in the integral in \eq{diff} putting $J_0 (0) =
1$. Integration over $z_T$ then yields
\ben
 \frac{C_F\, S_p\,
  S_A}{\as \pi (2\pi)^2} \, \int \, \frac{d\lambda}{2\pi i} \,
  \int\,\frac{d\lambda'}{2\pi i}\, C_{\lambda'}^p \, \lambda^2 \, \lambda'^2
  \, e^{2 \, \bas \, (Y-y) \, \chi(\lambda')}
\een
\be
\times \, \frac{1}{\lambda+\lambda'-2} \, \left(\frac{\Lambda}
{k_{\rm geom}}\right)^{\lambda'} \,
\frac{k_{\rm geom}^2}{Q_s^2 (y)} \, \bigg[ C_\lambda^A \,
e^{2 \, \bas \, y\, \chi(\lambda)} \, \left(\frac{Q_{s0}}{k_{\rm
geom}}\right)^\lambda - \tilde C_\lambda^A \, \left(\frac{Q_s (y)}
{k_{\rm geom}}\right)^\lambda\bigg]. \label{app1}
\ee 
Due to the inequality $k_{\rm geom} \gg Q_s(y) \gg \Lambda$, the
integration over $\lambda'$ in \eq{app1} is dominated by the saddle
point at $\lambda' \approx 2$, as shown in Eqs. (\ref{spd}) and
(\ref{spd2}).  The integral over $\lambda$ in \eq{app1} becomes
\begin{eqnarray}\label{app2}
&&\int \frac{d\lambda}{2\pi i} \, \frac{\lambda^2}{\lambda+\lambda'-2} \,
\left[C_\lambda^A\, e^{2\, \bas \, y \, \chi(y)}\,
\left(\frac{Q_{s0}}{k_{\rm geom}}\right)^\lambda\,-\,
\tilde C_\lambda^A\, \left(\frac{Q_{s}(y)}{k_{\rm geom}}\right)^\lambda
\right]\nonumber\\
&\approx & \int \frac{d\lambda}{2\pi i} \, \lambda \, 
\left[C_\lambda^A\, e^{2 \, \bas \, y \, \chi(y)}\,
\left(\frac{Q_{s0}}{k_{\rm geom}}\right)^\lambda\,-\,
\tilde C_\lambda^A\, \left(\frac{Q_{s}(y)}{k_{\rm geom}}\right)^\lambda
\right]\nonumber\\
&=& - \frac{\partial}{\partial \ln k_{\rm geom}}\,\left[ N_G (z_T
\rightarrow (1/k_{\rm geom})^-,y)\,-\,N_G (z_T \rightarrow (1/k_{\rm
geom})^+,y)\right]\,=\,0 ,
\end{eqnarray}
where we assumed that $N_G (\un z,y)$ and its derivatives with respect
to $z_T$ from \eq{exmel} are smooth functions of $z_T$ such that the
difference of the above limits is zero. This assumption is justified
since $N_G (\un z,y)$ is proportional to the scattering matrix which
is an analytic function of its variables. \eq{eqN} makes $N_G (\un
z,y)$ analytic by construction. 

We showed that the difference between the exact \eq{uwa} and our
\eq{f1} is zero, making the two equations equal, as desired. However, the 
above proof required that the representation of $N_G (\un z,y)$ given
by \eq{exmel} has a smooth matching of the two regions at $z_T =
1/k_{\rm geom}$, i.e., that representation (\ref{exmel}) is not just a
good approximation but an exact identity. To show that no such
assumption is required to prove that the expression in
\eq{app1} is a negligible correction to \eq{f1} let us estimate the energy
dependence the first term in \eq{app1}. The second term in \eq{app1}
is negative and can only make the overall contribution smaller.
Employing double logarithmic approximation for $\lambda$- and
$\lambda'$-integrals and using Eqs. (\ref{saty}) and (\ref{geom}) we
derive (setting $y=Y/2$ for simplicity)
\ben
\int\, \frac{d\lambda}{2\pi i} \int\,\frac{d\lambda'}{2\pi i} \, 
 C_{\lambda'}^p \, \lambda^2 \, \lambda'^2 \, 
 e^{2 \, \bas \, (Y-y) \,\chi(\lambda')} \,
 \frac{1}{\lambda+\lambda'-2} \, \left(\frac{\Lambda}
{k_{\rm geom}}\right)^{\lambda'}
\frac{k_{\rm geom}^2}{Q_s^2 (y)}\,C_\lambda^A\,
e^{2 \, \bas \, y \, \chi(\lambda)}\left(\frac{Q_{s0}}{k_{\rm
      geom}}\right)^\lambda
\een
\be\label{1st}
\propto \, \frac{\Lambda^2\, Q_{s0}^2 \, A^{1/6\sqrt{2}}}
{k_{\rm geom}^2 \, Q_s^2 (y)}\,
e^{8 \, \sqrt{2} \, \bas \, y}
\, \propto \, A^{-\frac{1}{3} + \frac{1}{6 \sqrt{2}}} \, 
e^{ - 4 \, \bas \, y (3 - 2 \sqrt{2})},
\ee
which is a decreasing function of rapidity and centrality. It is
obviously negligible compared to the increasing function of $y$ given
by \eq{f2}. This accomplishes our proof of \eq{f1}.

\end{document}